\documentclass[preprint2]{aastex}

\newcommand{\ts}{\thinspace}

\shorttitle{COSMOS-30 Photometric Redshifts}
\shortauthors{}

\begin{document}


\title{COSMOS Photometric Redshifts with 30-bands for 2-deg$^2$ \altaffilmark{1}}


\author{
O. Ilbert\altaffilmark{2},
P. Capak\altaffilmark{3,4},
M. Salvato \altaffilmark{3},
H. Aussel\altaffilmark{5},  
H. J. McCracken\altaffilmark{6},
D. B. Sanders\altaffilmark{2},
N. Scoville\altaffilmark{3}, 
J. Kartaltepe\altaffilmark{2},
S. Arnouts\altaffilmark{7},
E. Le Floc'h\altaffilmark{2},
B. Mobasher\altaffilmark{8},
Y. Taniguchi\altaffilmark{9},
F. Lamareille\altaffilmark{10},
A. Leauthaud\altaffilmark{11},
S. Sasaki\altaffilmark{12,13},
D. Thompson\altaffilmark{3,14},
M. Zamojski\altaffilmark{3},
G. Zamorani\altaffilmark{15},
S. Bardelli\altaffilmark{15},
M. Bolzonella\altaffilmark{15},
A. Bongiorno\altaffilmark{11},
M. Brusa\altaffilmark{16},
K.I. Caputi\altaffilmark{17},
C.M. Carollo\altaffilmark{17},
T. Contini\altaffilmark{10},
R. Cook\altaffilmark{3},
G. Coppa\altaffilmark{15},
O. Cucciati\altaffilmark{18},
S. de la Torre\altaffilmark{12},
L. de Ravel\altaffilmark{12},
P. Franzetti\altaffilmark{19},
B. Garilli\altaffilmark{19},
G. Hasinger\altaffilmark{16},
A. Iovino\altaffilmark{18},
P. Kampczyk\altaffilmark{17},
J.-P. Kneib\altaffilmark{11},
C. Knobel\altaffilmark{17},
K. Kovac\altaffilmark{17},
J.F. Le Borgne\altaffilmark{10},
V. Le Brun\altaffilmark{11},
O. Le F\`evre\altaffilmark{12}, 
S. Lilly\altaffilmark{17},
D. Looper\altaffilmark{2},
C. Maier\altaffilmark{17}, 
V. Mainieri\altaffilmark{10},
Y. Mellier\altaffilmark{6},
M. Mignoli\altaffilmark{15}, 
T. Murayama,    \altaffilmark{13}
R. Pell\`o\altaffilmark{10},
Y. Peng\altaffilmark{17},
E. P\'erez-Montero\altaffilmark{10},
A. Renzini\altaffilmark{20}
E. Ricciardelli\altaffilmark{20},
D. Schiminovich\altaffilmark{21},
M. Scodeggio\altaffilmark{19},
Y. Shioya\altaffilmark{9},
J. Silverman\altaffilmark{17}, 
J. Surace\altaffilmark{4}, 
M. Tanaka\altaffilmark{22},
L. Tasca\altaffilmark{11}, 
L. Tresse\altaffilmark{11},
D. Vergani\altaffilmark{19},
E. Zucca\altaffilmark{15}
}

\email{}



\begin{abstract}

We present accurate photometric redshifts in the 2-deg$^2$ COSMOS
field. The redshifts are computed with 30 broad, intermediate, and
narrow bands covering the UV ({\it GALEX}), Visible-NIR (Subaru, CFHT,
UKIRT and NOAO) and mid-IR ({\it Spitzer}/IRAC). A $\chi^2$
template-fitting method ({\it Le Phare}) was used and calibrated with
large spectroscopic samples from VLT-VIMOS and Keck-DEIMOS. We develop
and implement a new method which accounts for the contributions from
emission lines ([\ion{O}{2}], H$\beta$, H$\alpha$ and Ly$\alpha$) to
the spectral energy distributions (SEDs). The treatment of emission
lines improves the photo-z accuracy by a factor of 2.5. Comparison of
the derived photo-z with 4148 spectroscopic redshifts (i.e. $\Delta z
= z_{\rm s} - z_{\rm p}$) indicates a dispersion of $\sigma_{\Delta
z/(1+z_{\rm s})}=0.007$ at $i^+_{\rm AB}<22.5$, a factor of $2-6$
times more accurate than earlier photo-z in the COSMOS, CFHTLS and
COMBO-17 survey fields. At fainter magnitudes $i^+_{\rm AB}<24$ and
$z<1.25$, the accuracy is $\sigma_{\Delta z/(1+z_{\rm s})}=0.012$.
The deep NIR and IRAC coverage enables the photo-z to be extended to
$z\sim2$ albeit with a lower accuracy ($\sigma_{\Delta z/(1+z_{\rm
s})}=0.06$ at $i^+_{\rm AB}\sim 24$). The redshift distribution of
large magnitude-selected samples is derived and the median redshift is
found to range from $z_{\rm m}=0.66$ at $22 <i^+_{\rm AB}<22.5$ to
$z_{\rm m}=1.06$ at $24.5 <i^+_{\rm AB}<25$. At $i^+_{\rm AB} <26.0$,
the multi-wavelength COSMOS catalog includes approximately 607,617
objects. The COSMOS-30 photo-z enable the full exploitation of this
survey for studies of galaxy and large scale structure evolution at
high redshift.

\end{abstract}



\altaffiltext{1}{Based on observations with the NASA/ESA {\em
Hubble Space Telescope}, obtained at the Space Telescope Science
Institute, which is operated by AURA Inc, under NASA contract NAS
5-26555. Also based on observations made with the Spitzer Space Telescope,
which is operated by the Jet Propulsion Laboratory, California Institute
of Technology, under NASA contract 1407. Also based on data collected at :
the Subaru Telescope, which is operated by the National Astronomical 
Observatory of Japan; the XMM-Newton, an ESA science mission with
instruments and contributions directly funded by ESA Member States and
NASA; the European Southern Observatory under Large Program 175.A-0839,
Chile; Kitt Peak National Observatory, Cerro Tololo Inter-American
Observatory and the National Optical Astronomy Observatory, which are
operated by the Association of Universities for Research in Astronomy, Inc.
(AURA) under cooperative agreement with the National Science Foundation;
and the Canada-France-Hawaii Telescope with MegaPrime/MegaCam operated as a
joint project by the CFHT Corporation, CEA/DAPNIA, the NRC and CADC of
Canada, the CNRS of France, TERAPIX and the Univ. of Hawaii.}
\altaffiltext{2}{Institute for Astronomy, 2680 Woodlawn Dr., University of Hawaii, Honolulu, Hawaii, 96822}
\altaffiltext{3}{California Institute of Technology, MC 105-24, 1200 East California Boulevard, Pasadena, CA 91125}
\altaffiltext{4}{Spitzer Science Center, California Institute of Technology, Pasadena, CA 91125}
\altaffiltext{5}{AIM Unit\'e Mixte de Recherche CEA  CNRS Universit\'e Paris VII UMR n158}
\altaffiltext{6}{Institut d'Astrophysique de Paris, UMR7095 CNRS, Universit\'e Pierre et Marie Curie, 98 bis Boulevard Arago, 75014 Paris, France}
\altaffiltext{7}{Canada France Hawaii telescope corporation, 65-1238 Mamalahoa Hwy, Kamuela, Hawaii 96743, USA}
\altaffiltext{8}{Department of Physics and Astronomy, University of California, Riverside, CA, 92521, USA}
\altaffiltext{9}{Research Center for Space and Cosmic Evolution, Ehime University, Bunkyo-cho, Matsuyama 790-8577, Japan}
\altaffiltext{10}{Max Planck Institut f\"{u}r Extraterrestrische Physik,  D-85478 Garching, Germany}
\altaffiltext{11}{Laboratoire d'Astrophysique de Toulouse/Tarbes, Universit\'e de Toulouse, CNRS, 14 avenue E. Belin, 31400 Toulouse, France }
\altaffiltext{12}{Laboratoire d'Astrophysique de Marseille, BP 8, Traverse du Siphon, 13376 Marseille Cedex 12, France}
\altaffiltext{13}{Astronomical Institute, Graduate School of Science, Tohoku University, Aramaki, Aoba, Sendai 980-8578, Japan}
\altaffiltext{14}{Physics Department, Graduate School of Science \& Engineering, Ehime University, 2-5 Bunkyo-cho, Matsuyama, 790-8577, Japan}       
\altaffiltext{15}{LBT Observatory, University of Arizona, 933 N. Cherry Ave., Tucson, Arizona, 85721-0065, USA}
\altaffiltext{16}{INAF-Osservatorio Astronomico di Bologna, via Ranzani 1, I-40127 Bologna, Italy}
\altaffiltext{17}{Department of Physics, ETH Zurich, CH-8093 Zurich, Switzerland}
\altaffiltext{18}{INAF Osservatorio Astronomico di Brera, Milano, Italy}
\altaffiltext{19}{INAF - IASF Milano, via Bassini 15, 20133 Milano, Italy}
\altaffiltext{20}{Dipartimento di Astronomia, Universitˆ di Padova, vicolo dell'Osservatorio 2, I-35122 Padua, Italy}
\altaffiltext{21}{Department of Astronomy, Columbia University, MC2457, 550 W. 120 St. New York, NY 10027}
\altaffiltext{22}{European Southern Observatory, Karl-Schwarzschild-Str. 2, D-85748 Garching, Germany}

\section{Introduction}

Photometric redshifts (hereafter photo-z) are an estimate of galaxy
distances based on the observed colors (Baum 1962). This method is
extremely efficient for assembling large redshift samples for faint
galaxies. Despite having a lower accuracy than spectroscopic redshifts
(hereafter, spectro-z), photo-z have the advantage of significantly
improved completeness down to a flux limit fainter than the
spectroscopic limit.  Deep photo-z samples such as COMBO-17 (Wolf et
al. 2003), CFHTLS (Ilbert et al. 2006), SWIRE (Rowan-Robinson et
al. 2008), and COSMOS (Mobasher et al. 2007) contain more than
1,000,000 galaxies and go as faint as $i\sim 25$ with a relatively
small amount of telescope time.

Typical photo-z with an accuracy of $\sigma_{\Delta z/(1+z_{\rm s})}
\sim0.02-0.04$ ($\Delta z = z_{\rm s}-z_{\rm p}$) are widely
used to study the evolution of galaxy stellar masses and luminosities
(e.g. Fontana et al. 2000; Wolf et al. 2003; Gabasch et al. 2004;
Caputi et al. 2006; Arnouts et al. 2007), for angular clustering
analysis (e.g. Heinis et al. 2007; McCracken et al. 2007), to study
the relation between galaxy properties and environment (e.g. Capak et
al. 2007), to trace large-scale structures (Mazure et al. 2007;
Scoville et al. 2007) and to identify clusters at high redshift (Wang
\& Steinhardt 1998). Photo-z are also necessary for dark energy and
dark matter weak lensing studies to separate foreground and background
galaxies and to control systematic effects such as intrinsic shape
alignment, shear shape correlation, and the effects of source
clustering (Peacock et al. 2006). All of these applications require
strict control of systematic effects in the photo-z estimate.  An
efficient way of identifying and removing systematics is to calibrate
photo-z on a spectroscopic sample. The most common methods of
calibration are neural network methods (e.g. Ball et al. 2004,
Colister et al. 2004), a calibration of the color-z relation (Brodwin
et al. 2006; Ilbert et al. 2006), or a reconstruction of the SED
templates (Budav\'ari et al. 2000; Feldmann et al. 2006).

As is also true for spectroscopic redshift measurements, photo-z
accuracy depends on spectral coverage and resolution. It is also
degraded for sources with a low signal-to-noise ratio (Bolzonella et
al. 2000).  The Balmer and Lyman breaks contain much of the photo-z
information, so accuracy is lower in redshift ranges where these
features are not well sampled by the filter set. As the photometric
accuracy degrades, it becomes more difficult to constrain the
positions of these features, leading to lower accuracy.  This creates
a dual dependency of photo-z accuracy on magnitude and redshift which
is defined by the survey design (exposure time, filter choice, and
calibration accuracy). For this reason, any photo-z sample should be
extensively tested and characterized in the same way a spectroscopic
sample would be. This analysis is necessary in order to identify the
redshift/magnitude ranges over which the photo-z can be trusted and
used for a given scientific application.

This paper presents a new version of the photometric redshifts for the
Cosmic Evolution Survey (COSMOS: Scoville et al. 2007) and an analysis
of their accuracy. COSMOS is the largest {\it Hubble Space Telescope
(HST)} survey ever undertaken - imaging an equatorial 2-deg$^2$ field
to a depth of $I_{\rm F814W} = 27.8$ mag (5 $\sigma$, AB). The COSMOS
field is equatorial to ensure visibility by all ground-based
astronomical facilities. This project includes extensive
multi-wavelength imaging from X-ray to radio ({\it XMM}, {\it
Chandra}, {\it GALEX}, Subaru, CFHT, UKIRT, {\it Spitzer}, VLA).  New
ground-based NIR data (McCracken et al. 2008; Capak et al. 2008), {\it
Spitzer}-IRAC data (Sanders et al. 2007), and medium/narrow band data
from the Subaru Telescope (Taniguchi et al. 2007; Sasaki et al. 2008;
Taniguchi et al. 2008; Capak et al. 2008) greatly improve the previous
photometry catalogue (Capak et al. 2007). These new data are used for
the photo-z derived here, yielding a factor of 3 higher accuracy than
the first release of COSMOS photo-z (Mobasher et al. 2007).

The COSMOS data are presented in \S2. The technique used to estimate
the photo-z is presented in \S3. In \S4, we quantify the photo-z
accuracy as a function of the magnitude and redshift. In \S5, we
provide the photo-z distribution of the $i^+$ selected
samples. Specialized photo-z for X-ray selected sources, AGN, and
variable objects, are discussed in a companion paper (Salvato et
al. 2008).

Throughout this paper, we use the standard WMAP cosmology
($\Omega_{\rm m}~=~0.3$, $\Omega_\Lambda~=~0.7$) with $h=0.7$ and
$h~=~H_{\rm0}/100$~km~s$^{-1}$~Mpc$^{-1}$.  Magnitudes are given in the
AB system.

\section{Data}

Compared to the previous optical/near -infrared catalogue (Capak
et al. 2007) the new photometry implements 14 new medium/narrow band
data from the Subaru Telescope, deep ground-based NIR data ($J$ and
$K$ bands) and {\it Spitzer}-IRAC data. The spectroscopic sample used
to calibrate/test the photo-z is 10 times larger at $i^+_{AB}<22.5$
than that of Mobasher et al. (2007). The spectroscopic sample is
supplemented with faint infrared selected sources and a deep faint
spectroscopic sample at $z>1.5$. Hereafter, we detail the photometric
and spectroscopic data used to measure the photo-z. 

\subsection{Photometric data}

Fluxes are measured in 30 bands from data taken on the Subaru
(4200-9000\AA), CFHT (3900-21500\AA), UKIRT (12500\AA), {\it Spitzer}
(3.6-8$\mu m$) and {\it GALEX} (1500-2300\AA) telescopes. We refer to
Capak et al. (2008) for a complete description of the observations,
data reduction and the photometry catalogue. The equivalent width and
the effective wavelength of each filter are listed in
Table~\ref{shift}. The sensitivities are given in Capak et al. (2008)
and in Table~1 of Salvato et al. (2008).  Table~\ref{shift} indicates
the fraction of sources detected in each band with an error less than
0.2{\ts}mag for a selection at $i^+ <24.5$, $i^+ <25$ and $i^+
<25.5$. We summarize below the datasets used in this paper.

{\bf Ultraviolet:} Very deep $u^*$ band data were obtained at the 3.6m
Canada-France Hawaii Telescope (CFHT) using the Megacam camera
(Boulade et al. 2003).  The $u^*$ band data were processed at the
TERAPIX data reduction center\footnote{terapix.iap.fr}. The $u^*$ band
data cover the entire COSMOS field and reach a depth of $u^* \sim
26.5${\ts}mag for a point source detected at 5$\sigma$. The $u^*$ band
images are also used as priors in the measurement of FUV (1500\AA) and
NUV (2300\AA) fluxes in order to ensure a proper deblending of sources
in the GALEX images (Zamojski et al. 2007). GALEX fluxes are then
extracted using the EM-algorithm (Guillaume et al. 2006).  They reach
a depth of $FUV \sim 26$ mag and $NUV \sim 25.7$ mag.

{\bf Optical:} The COSMOS-20 survey (Taniguchi et al. 2008) entailed
30 nights of observation at the Subaru 8.2m telescope using the
Suprime-Cam instrument. The observations are complete in 20 bands: 6
broad bands ($B_J$, $V_J$, $g^+$, $r^+$, $i^+$, $z^+$), 12 medium
bands ($IA427$, $IA464$, $IA484$, $IA505$, $IA527$, $IA574$, $IA624$,
$IA679$, $IA709$, $IA738$, $IA767$, $IA827$) and 2 narrow bands
($NB711$, $NB816$).

{\bf Near-infrared:} The catalogue includes deep $J$ and $K$
band data obtained using the WFCAM and WIRCAM wide-field infrared
cameras on UKIRT and CFHT, respectively.  The NIR data reduction is
detailed in Capak et al. (2008) and McCracken et al. (2008).  The data
reach $J \sim 23.7${\ts}mag and $K \sim 23.7${\ts}mag for a
point source detected at 5$\sigma$.

{\bf Mid-infrared:} Deep IRAC data were taken during the {\it Spitzer}
cycle 2 S-COSMOS survey (Sanders et al. 2007).  A total of 166{\ts}hr
were dedicated to cover the full 2-deg$^2$ with the IRAC camera in 4
bands: 3.6$\mu$m, 4.5$\mu$m, 5.6$\mu$m and 8.0$\mu$m. Source detection
is based on the 3.6$\mu m$ image and the fluxes were measured in the
four IRAC bands using the ``dual mode'' configuration of
SExtractor. The IRAC catalogue is 50\% complete at $1\mu Jy$ at
$3.6\mu m$ ($m_{3.6\mu m} \sim 23.9$ mag).

All of the imaging data were combined to generate a master photometry
catalogue (Capak et al. 2008).  Photometry was done using SExtractor
in dual mode (Bertin \& Arnouts 1996). Source detection was run on the
deepest image ($i^+ \sim 26.2$ for a point source detected at
5$\sigma$).  For the UV-NIR data, the Point Spread Function (PSF)
varies from 0.5\arcsec\ to 1.5\arcsec\ from the $K$ to the $u*$
images. In order to obtain accurate colors, all the images were
degraded to the same PSF of 1.5\arcsec\ following the method described
in Capak et al. (2007). The final photometry catalogue contains PSF
matched photometry for all the bands from the $u^*$ to the $K$ band,
measured over an aperture of 3\arcsec\ diameter at the position of the
$i^+$ band detection. For the FUV and NUV data, we transformed the
total flux provided for the {\it GALEX}-FUV and -NUV counterpart by
multiplying it by a factor 0.759. This factor is the fraction of the
flux observed in optical into a 3\arcsec\ aperture flux as determined
for point sources from simulations by Capak et al. (2007). To
compensate for this approximation, we add in quadrature 0.1 and
0.3{\ts}mag to the all the measured errors in the GALEX NUV and FUV
bands.

For the MIR data we did not degrade the optical data to the larger IRAC PSF.  
Instead, the following procedures were used.  An IRAC flux 
was first associated with the optical sources by matching their positions 
in each of the 4 IRAC bands within a search radius of 1\arcsec. 
Following Surace et al. (2004), the IRAC fluxes were then measured 
in a circular aperture of radius 1.9\arcsec.
The aperture flux was then converted to a total flux using the
aperture correction factors 0.76, 0.74, 0.62, 0.58 at 3.6$\mu m$,
4.5$\mu$m, 5.6$\mu$m and 8.0$\mu$m, respectively (Surace et
al. 2004). Since the optical fluxes were measured over an aperture of
3\arcsec\ diameter which encloses $\sim 75\%$ of the flux for a point
like source, we then multiply the IRAC total fluxes by a factor
0.75. This approximation provides good agreement between the predicted
and observed colors ($z^+ - 3.6\mu m$ and $3.6\mu m - 4.5\mu m$
colors).  In order to compensate for this approximation, we add in
quadrature 0.1, 0.1, 0.3 and 0.3{\ts}mag to the errors in the IRAC
3.6, 4.5, 5.6 and 8.0$\mu$m. 

Finally, all magnitudes are corrected for galactic extinction
estimated for each object individually, using dust map images from
Schlegel et al. (1998). We limit the photo-z analysis to an area of
2-deg$^2$ ($1.49878<\alpha<2.91276$ and $149.41140<\delta<150.826934$)
which has a uniform and deep coverage in all the bands. Poor image
quality areas (e.g. field boundary, saturated stars, satellite tracks
and image defects) are masked. Photo-z are computed only in the
non-masked regions with a total covered area of 1.73-deg$^2$. 126 071,
293 627 and 607 617 sources are detected at $i^+<24$, $i^+<25$ and
$i^+<26$.

\begin{table*}
\begin{center}
\begin{tabular}{c c c c c c c c c } \hline \\

                 filter & telescope& effective $\lambda$ & FWHM & $s_f$ & \% at &
                 \% at & \% at \\ & & & & & $i^+ <24.5$ & $i^+ <25$ &
                 $i+ <25.5$ \\
\\
\hline\\
                  $u*$  & CFHT           &  3911.0   &     538.0 &   0.054  &    89.3  &  85.2  &  77.2  \\ 
           $B_{\rm J}$  & Subaru         &  4439.6   &     806.7 &  -0.242  &    97.1  &  95.2  &  90.5	\\ 
           $V_{\rm J}$  & Subaru         &  5448.9   &     934.8 &  -0.094  &    99.3  &  98.2  &  94.2	\\ 
                $g^+$   & Subaru         &  4728.3   &    1162.9 &   0.024  &    96.4  &  93.6  &  86.0	\\ 
                $r^+$   & Subaru         &  6231.8   &    1348.8 &   0.003  &    99.6  &  99.5  &  98.4	\\ 
                $i^+$   & Subaru         &  7629.1   &    1489.4 &   0.019  &    99.9  &  99.9  &  99.8	\\ 
                $i*$    & CFHT           &  7628.9   &    1460.0 &  -0.007  &    37.8  &  25.4  &  17.4 \\ 
                $z^+$   & Subaru         &  9021.6   &     955.3 &  -0.037  &    99.8  &  97.9  &  83.8	\\ 
                $J$     & UKIRT          &  12444.1  &    1558.0 &   0.124  &    65.4  &  49.1  &  35.7	\\ 
                 $K_S$  & NOAO           &  21434.8  &    3115.0 &   0.022  &    15.3  &  10.3  &  7.08	\\ 
                   $K$  & CFHT           &  21480.2  &    3250.0 &  -0.051  &    84.1  &  68.5  &  52.1	\\ 
               $IA427$  & Subaru         &  4256.3   &     206.5 &   0.037  &    77.1  &  64.3  &  48.4	\\ 
               $IA464$  & Subaru         &  4633.3   &     218.0 &   0.013  &    78.5  &  64.3  &  47.6	\\ 
               $IA484$  & Subaru         &  4845.9   &     228.5 &   0.000  &    88.7  &  78.3  &  62.0	\\ 
               $IA505$  & Subaru         &  5060.7   &     230.5 &  -0.002  &    84.0  &  70.0  &  52.0	\\ 
               $IA527$  & Subaru         &  5258.9   &     242.0 &   0.026  &    93.2  &  84.6  &  68.7	\\ 
               $IA574$  & Subaru         &  5762.1   &     271.5 &   0.078  &    92.5  &  80.0  &  60.6	\\ 
               $IA624$  & Subaru         &  6230.0   &     300.5 &   0.002  &    97.4  &  90.2  &  72.2	\\ 
               $IA679$  & Subaru         &  6778.8   &     336.0 &  -0.181  &    99.5  &  96.7  &  82.9	\\ 
               $IA709$  & Subaru         &  7070.7   &     315.5 &  -0.024  &    99.8  &  97.7  &  83.5	\\ 
               $IA738$  & Subaru         &  7358.7   &     323.5 &   0.017  &    99.5  &  94.2  &  73.6	\\ 
               $IA767$  & Subaru         &  7681.2   &     364.0 &   0.041  &    99.6  &  93.5  &  72.0	\\ 
               $IA827$  & Subaru         &  8240.9   &     343.5 &  -0.019  &    99.7  &  97.2  &  81.8	\\ 
               $NB711$  & Subaru         &  7119.6   &      72.5 &   0.014  &    84.8  &  60.7  &  41.9	\\ 
               $NB816$  & Subaru         &  8149.0   &     119.5 &   0.068  &    99.8  &  99.1  &  88.2	\\ 
               $IRAC1$  & {\it Spitzer} &  35262.5  &    7412.0 &   0.002  &    70.6  &  60.8  &  48.4	\\ 
               $IRAC2$  & {\it Spitzer} &  44606.7  &   10113.0 &   0.000  &    62.6  &  51.6  &  39.7	\\ 
               $IRAC3$  & {\it Spitzer} &  56764.4  &   13499.0 &   0.013  &    33.7  &  26.0  &  19.5	\\ 
               $IRAC4$  & {\it Spitzer} &  77030.1  &   28397.0 &  -0.171  &    15.7  &  11.3  &   8.1 	\\ 
                 $FUV$  &  {\it GALEX}  &  1551.3   &     230.8 &   0.314  &     8.5  &   5.8  &   4.0 	\\ 
                 $NUV$  &  {\it GALEX}  &  2306.5   &     789.1 &  -0.022  &    19.7  &  13.4  &   9.2 	\\ 
\hline
\end{tabular}
\caption{Effective wavelength, width and systematic offsets $s_f$ in magnitude (with our definition, $s_f$ have the opposite sign to Table.13 of Capak et al. 2007) plus the fraction of sources in each band with an error less than 0.2 magnitude at $i^+ <24.5$, $i^+ <25$, $i^+ <25.5$. }
\label{shift}
\end{center}
\end{table*}

\subsection{Spectroscopic data}

The spectroscopic samples were observed with the VIMOS/VLT
spectrograph (zCOSMOS; Lilly et al. 2007) and the DEIMOS/Keck
spectrograph (Kartaltepe et al. 2008). These two spectroscopic samples
have very different selection criteria (see below); they therefore
cover very different ranges of redshift and color space, providing a
broad sample for evaluation of the photo-z.

The zCOSMOS survey (Lilly et al. 2007) has two components:
zCOSMOS-bright with a sample of 20,000 galaxies selected at $i^* \le
22.5$ and zCOSMOS-faint with approximately 10,000 galaxies
color-selected to lie in the redshift range $1.5 \lesssim z
\lesssim 3$. In the latter, galaxies are selected by color based
either on the $BzK$ criterion (Daddi et al. 2004) or the $UGR$ ``BM"
and ``BX" criterion of Steidel et al. (2004), and the magnitude cut
was $B_{\rm J} <24-25$ (depending on the color cut). zCOSMOS-bright
galaxies were observed using the red grism of VIMOS covering a
wavelength range $5500\rm{ \AA} <
\lambda < 9000\rm{ \AA}$ at a resolution of 600 (MR grism). For the zCOSMOS-faint sample,
observations were carried out with the blue grism of VIMOS ($3600\rm{
\AA} < \lambda < 6800\rm{ \AA}$) at a resolution of 200.

The zCOSMOS-bright survey is now $\sim50$\% complete. Here we make use
of only the extremely secure spectro-z measurements with a confidence
level greater than 99\% (class 3 and 4).  This secure zCOSMOS-bright
sample contains 4148 galaxies with a median redshift of $\sim
0.48$. The zCOSMOS-faint survey is in its early stages, and here we
use a preliminary sample of 148 galaxies with a median redshift of
$z_{\rm m} \sim 2.2$ and as faint as $i^+\sim 25$. This zCOSMOS-faint
spectroscopic sample is not fully representative of the average
population at $1.5<z<3$ due to the selection criteria (e.g. $B_{\rm J}
<24-25$).

The Keck{\ts}II spectroscopic follow-up of $24 \mu m$ selected sources
(Kartaltepe et al. 2008) is on-going and we refer to this sample as
MIPS-spectro-z. The DEIMOS spectra cover a wavelength range $4000\rm{
\AA} < \lambda < 9000\rm{ \AA}$ at a resolution of 600. This sample of
$24 \mu m$ selected galaxies contains 317 secure spectro-z (at least
two spectral features) with an average redshift of $z\sim 0.74$ and
apparent magnitude in the range $18< i^+< 25$.

For all of the spectroscopic samples used in this paper for testing
and verification of the photo-z, we include only secure spectro-z.
Therefore, the uncertainties in the spectro-z are neglected and the
spectro-z are used as a reference to assess the quality of the
photo-z.

\section{Photometric Redshift derivation}

Photometric redshifts were derived using the {\it Le Phare}
code\footnote{www.oamp.fr/people/arnouts/LE\_PHARE.html} (S. Arnouts
\& O. Ilbert) which is based on a $\chi^2$ template-fitting procedure. 
In the discussion below, we focus on the improvements introduced here
as compared to Ilbert et al. (2006) and the previous COSMOS photo-z
(Mobasher et al. 2007).

\subsection{Galaxy SED template library}

\begin{figure}[htb]
\includegraphics[width=8cm]{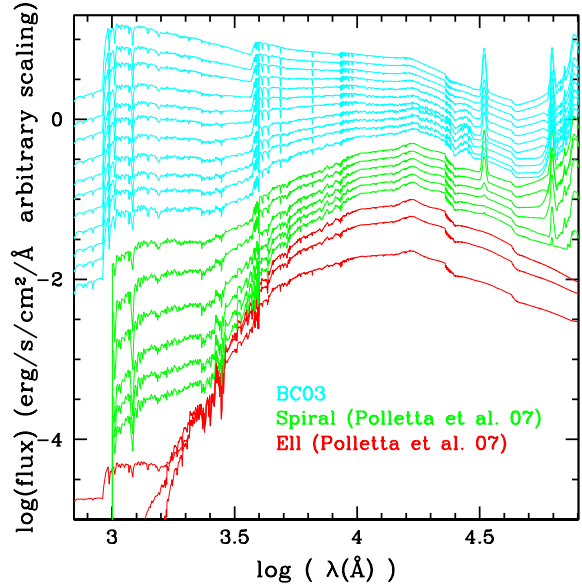}
\caption{SED templates. 
The flux scale is arbitrary. The top 12 SEDs (cyan) are generated with
Bruzual \& Charlot (2003).  The spiral (green) and elliptical (red)
SEDs are from Polletta et al. (2007). \label{SED}}
\end{figure}

Ilbert et al. (2006) and Mobasher et al. (2007) used a set of local
galaxy SED templates (CWW: Coleman, Wu, \& Weedman 1980) which have
been widely employed for photo-z (e.g. Sawicki 1997;
Fern{\'a}ndez-Soto et al. 1999; Arnouts et al. 1999; Brodwin et
al. 2006). Here, we employ a new set of templates generated by
Polletta et al. (2007) with the code GRASIL (Silva et al. 1998).
Polletta et al. selected their templates for fitting the VVDS sources
(Le F\`evre et al. 2005) from the UV-optical (CFHTLS: McCracken et
al. 2007) to the mid-IR (SWIRE: Lonsdale et al. 2003). Therefore, this
set of templates provides a better joining of UV and MIR than those by
CWW. The 9 galaxy templates of Polletta et al. (2007) include 3 SEDs of
elliptical galaxies and 6 templates of spiral galaxies (S0, Sa, Sb,
Sc, Sd, Sdm). 

We did find that the blue observed colors of the spectroscopic sample
were not fully reproduced by the Polletta et al. (2007) templates. We
therefore generated 12 additional templates using Bruzual \& Charlot
(2003) models with starburst (SB) ages ranging from 3 Gyr to 0.03 Gyr.
We extend the BC03 templates beyond $3\mu m$ rest-frame using the Sdm
template of Polletta et al. (2007). The full library of template SEDs,
9 from Polletta et al. (2007) and 12 from Bruzual \& Charlot (2003),
is shown in Fig.~\ref{SED}. Finally, we linearly interpolated between
some Polletta et al. templates to refine the sampling in $color-z$
space.

\begin{figure*}
\begin{tabular}{c c}
\includegraphics[width=8cm]{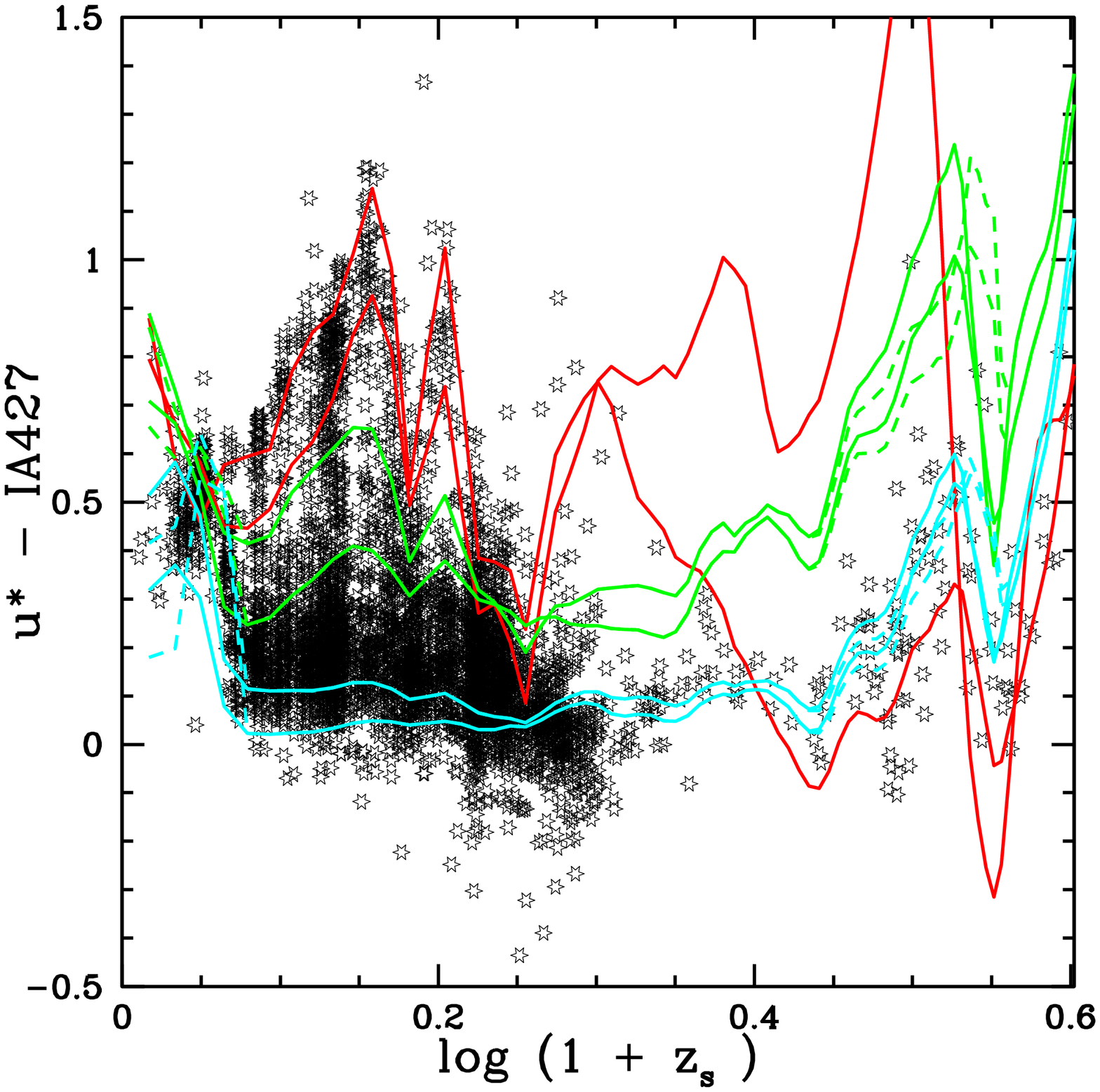}  &
\includegraphics[width=8cm]{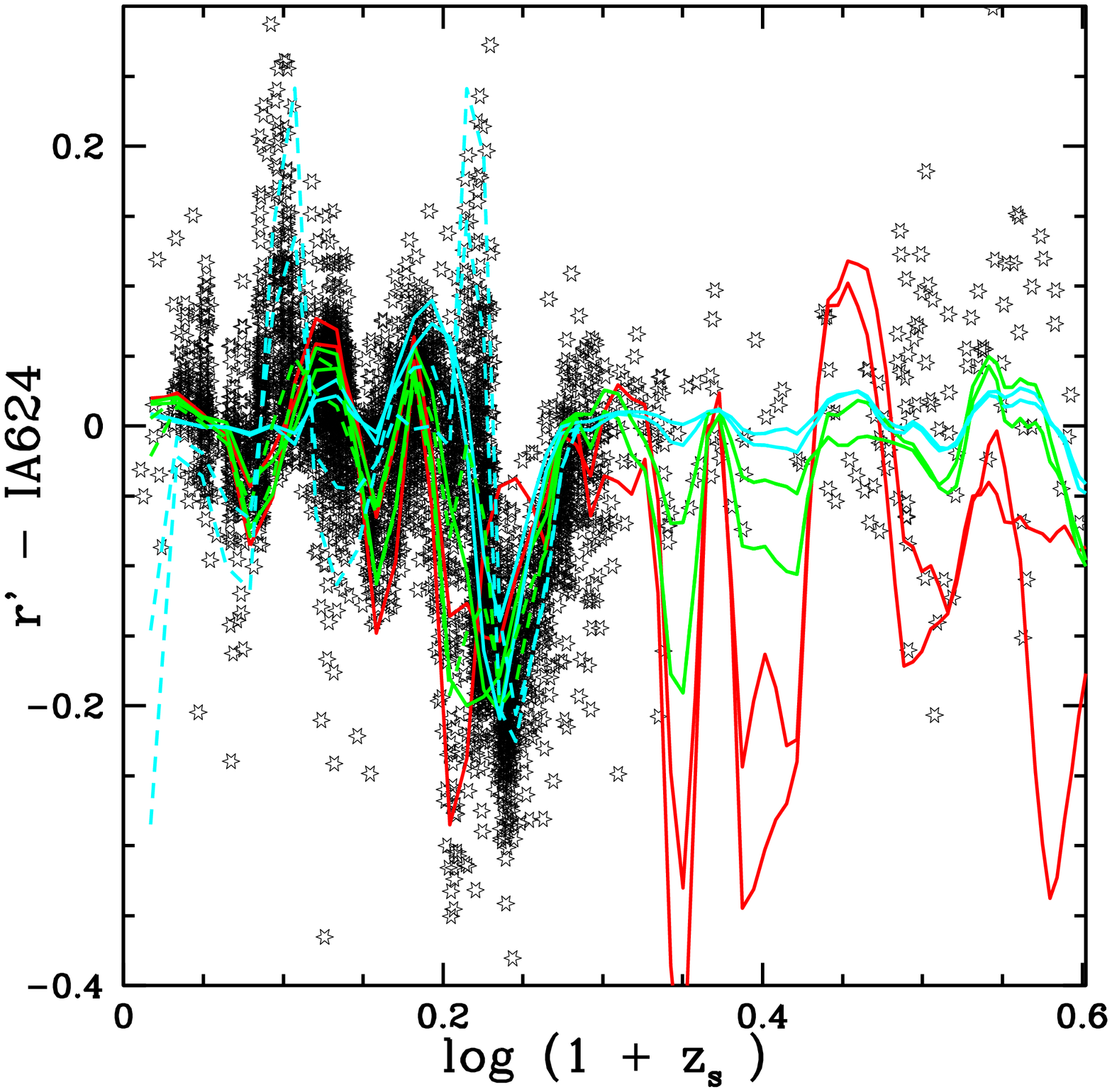} \\   
\includegraphics[width=8cm]{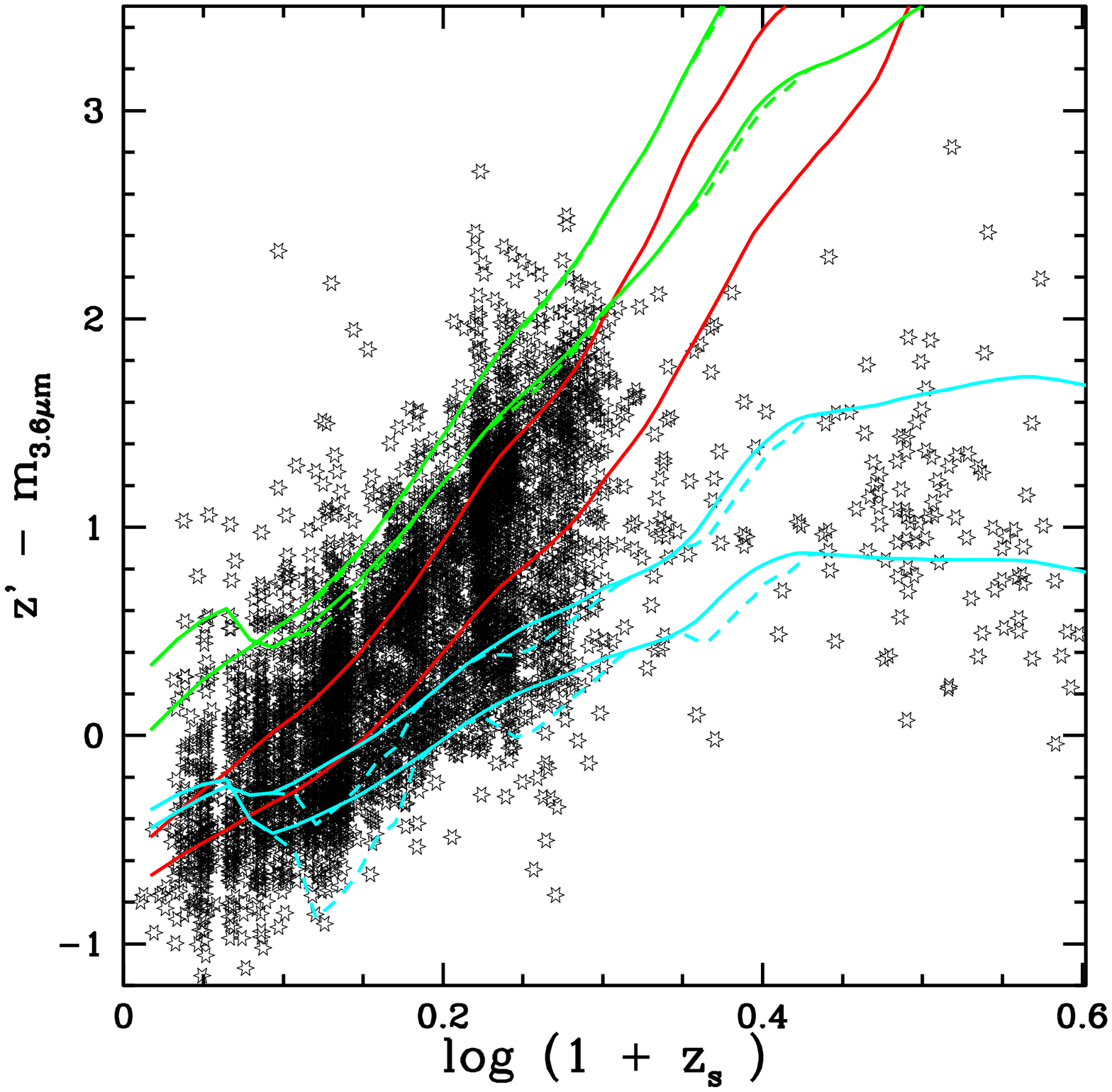}  &
\includegraphics[width=8cm]{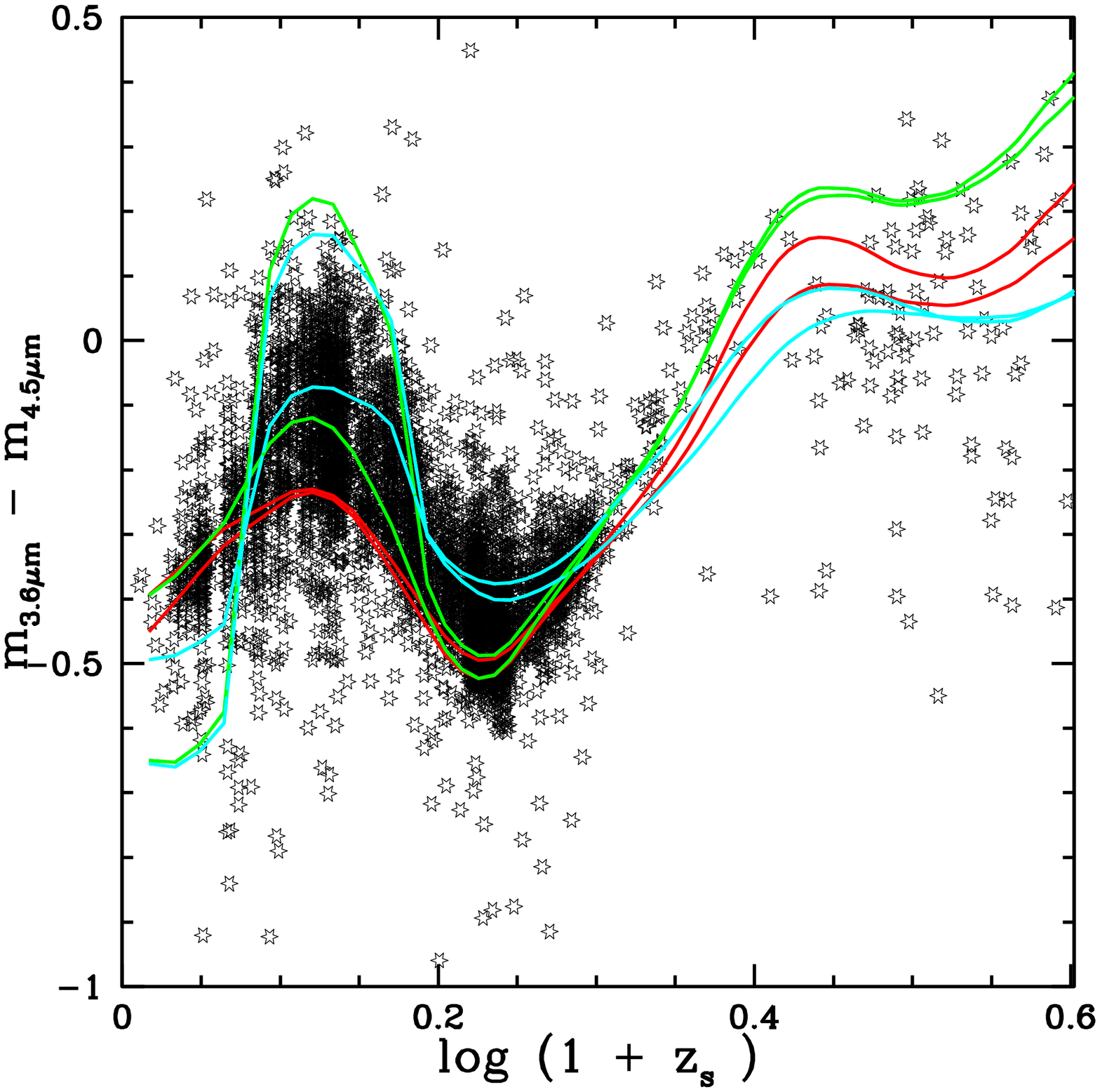} 
\end{tabular}
\caption{The observed colors and redshifts ($z_{\rm s}$) for the spectroscopic 
sample galaxies (open stars). The solid lines are the predicted colors
as a function of redshift for some SEDs of the library (red:
Elliptical and green: Spiral from Polletta et al. 2007 ; cyan: Bruzual
\& Charlot 2003). The solid curves are the predicted colors without
including emission lines (no reddening for the elliptical templates,
$E(B-V)=0.2$ for the late types) whereas the dashed curves are the
same templates including the emission line fluxes (assuming
$M_{FUV}=-20$ in this example). The top right panel clearly shows that
the emission lines can change the colors up to 0.4 mag.
\label{colour_z}}
\end{figure*}

Figure~\ref{colour_z} shows the observed colors and redshifts of the 
spectroscopic sample compared with the predicted colors for the library 
SEDs.

\subsection{Emission lines}

Fig.~\ref{colour_z} clearly shows how the observed colors oscillate
with the redshift, especially when the colors are measured with the
medium bands (top panels). Comparing the template curves with (dashed)
and without (solid) emission lines, one sees that the expected line
fluxes can cause up to 0.4{\ts}mag changes in the color. This effect
is particularly important when colors involving intermediate and
narrow band filters are computed (see for example the upper-right
panel in Fig.\ref{colour_z}) but can be already noticed using broad
band colors. The color oscillations are well explained by the
contribution of emission lines like H${\alpha}$, [\ion{O}{3}], and
[\ion{O}{2}] to the observed flux; thus the contribution of the
emission lines to the flux must be taken into account to obtain
accurate photo-z and this is a major change implemented here compared
to Ilbert et al. (2006) and Mobasher et al. (2007).

In order to include the emission line contribution to the SED, we need
to model the emission line fluxes (\ion{O}{2}, \ion{O}{3}, H$\beta$,
H$\alpha$, Ly$\alpha$) at any redshift, template and extinction. The
rescaling of the template ($A$ in eqn.\ref{chi2}) determines also the
emission line fluxes (therefore, the modeling of the fluxes must be
done galaxy per galaxy).

Our new procedure estimates the [\ion{O}{2}] emission line flux from
the UV luminosity of the rescaled template, using the Kennicutt (1998)
calibration laws. In the template fitting, a UV rest-frame luminosity
corrected for dust extinction can be computed at every step of the
redshift/template/extinction grid (the rescaling factor $A$ is taken
into account in the UV luminosity). The UV luminosity (at 2300\AA) is
then related to the SFR using the relation $SFR{\ts}(M_{\Sun}\
yr^{-1}) = 1.4 \times 10^{-28} L_{\nu}{\ts}(erg\ s^{-1} Hz^{-1}$) from
Kennicutt (1998). This SFR can then be translated to an [\ion{O}{2}]
emission line flux using the relation $SFR{\ts}(M_{\Sun}~yr^{-1})=
(1.4 \pm 0.4) \times 10^{-41} L_{\rm [OII]}{\ts}(erg~s^{-1})$
(Kennicutt 1998). This translates to :
\begin{equation}\label{FOII_eqn}
log (F_{\rm [OII]}) = -0.4\times M_{\rm UV} + 10.65 -
\frac{DM(z)}{2.5}
\end{equation}
where $DM$ is the distance modulus, $F_{\rm [OII]}$ is expressed in
units of $10^{-17}{\ts}erg\ s^{-1}cm^2$ and $M_{\rm UV}$ is the dust
corrected $UV$(2300\AA) absolute magnitude.

\begin{figure}[htb]
\includegraphics[width=8cm]{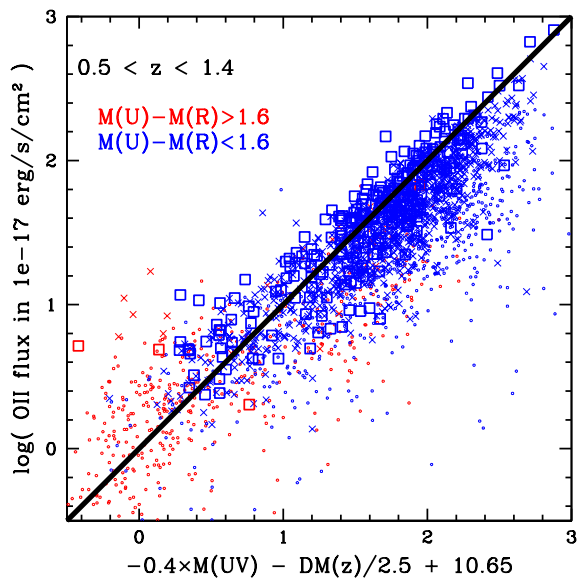}
\caption{Relation between the [\ion{O}{2}] flux and the absolute
magnitudes in UV (2300\AA). The solid line corresponds to the relation
obtained by applying the Kennicutt (1998) relations between $SFR_{\rm
OII}$ and $SFR_{\rm UV}$ as used here to include emission line fluxes
in the photo-z fitting (see eqn.\ref{FOII_eqn}) and the points are
observed emission line fluxes from VVDS (Lamareille et al. 2008).
(The UV luminosities and \ion{O}{2} fluxes are corrected for dust
extinction.) The red and blue points are galaxies with $M_U-M_R>1.6$
and $M_U-M_R<1.6$, respectively. The larger symbols correspond to
larger equivalent width.
\label{fOII_MNUV}}
\end{figure}

Figure~\ref{fOII_MNUV} shows the measured [\ion{O}{2}] fluxes from
VVDS (Lamareille et al. 2008) and the relation (solid line) expected
from Kennicutt (1998). We can perform this comparison only at
$0.5<z<1.4$, when the [\ion{O}{2}] line is observable in the VIMOS
spectra. Figure~\ref{fOII_MNUV} shows the good correlation between the
measured and predicted [\ion{O}{2}] fluxes, with an rms dispersion of
0.2 dex. In acknowledgment of this dispersion, we allow the intrinsic
[\ion{O}{2}] flux to vary by a factor of 2 in the templates with
respect to the nominal flux predicted by eqn.\ref{FOII_eqn}.

With our procedure, we can predict for each galaxy the [\ion{O}{2}]
flux at every redshift, template and extinction combination. For the
other emission lines, we adopt intrinsic, unextincted flux ratios of
[\ion{O}{3}/\ion{O}{2}] = 0.36; [H$\beta$/\ion{O}{2}] = 0.61;
[H$\alpha$/\ion{O}{2}] = 1.77 and [Ly$\alpha$/\ion{O}{2}] = 2 (McCall
et al. 1985, Moustakas et al. 2006, Mouhcine et al. 2005, Kennicutt
1998). When we apply an additional extinction to the template, we
modify these ratios with the corresponding attenuation. Then, we sum
the emission line fluxes to the template continuum before integrating
through the filter transmission curves.

The effect of these emission lines on the modeled $color-redshift$
relation is shown in Fig.~\ref{colour_z} for a galaxy at
$M_{UV}=-20$. The oscillations in the observed colors versus redshift
are well reproduced by the models.

\begin{figure*}
\includegraphics[width=15.5cm]{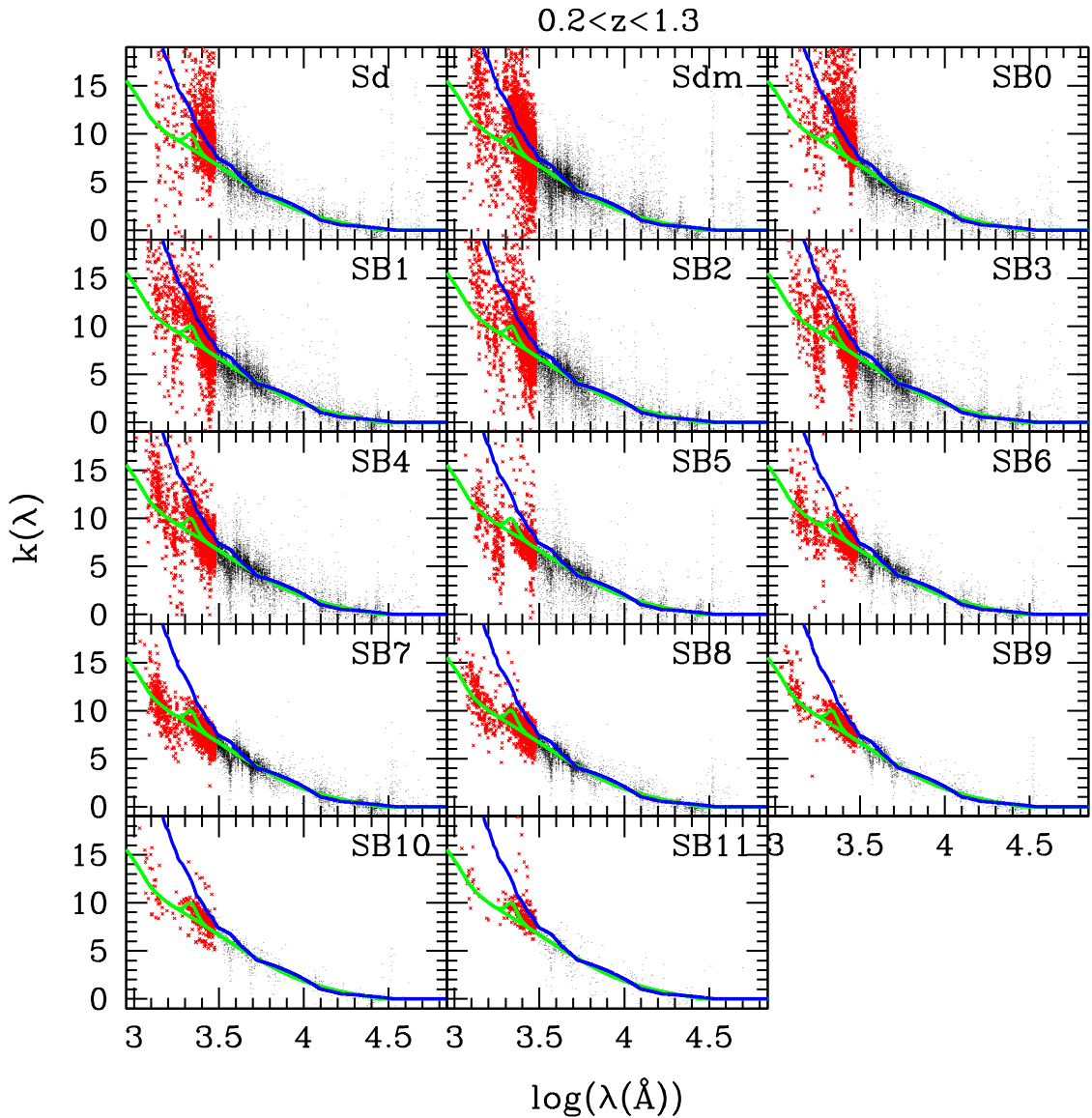}  
\caption{Attenuation by dust as a function of $\lambda$. The points 
are the extinction $A_i/(E_i(B-V))$ estimated from the galaxies with a
spectro-z (see \S3.4). The red points at $\lambda < 3000\rm{\AA}$ are
not used to estimate $E_i(B-V)$. The Prevot et al. (1984) and Calzetti
et al. (2000) extinction curves are shown with the blue and green
solid lines, respectively. The Prevot et al. (1984) extinction law is
rescaled to the same $A_{\rm V}$ as the Calzetti law by applying a
factor $4.05/2.72$. The extinction curve derived by Prevot et
al. (1984) is used for the galaxies redder than SB3 and the Calzetti
et al. (2000) extinction law for the galaxies bluer than
SB3.\label{SEDext}}
\end{figure*}

\subsection{Systematic offsets}

The $\chi^2$ template-fitting method is meaningful only if the
$color-z$ relation predicted from the templates is a good
representation of the observed $color-z$ relation. Uncertainties in
the zero-point offsets of photometric bands can lead to systematic
shifts between the predicted $color-z$ relation and the observed
colors of the spectroscopic sample.

To evaluate the zero-point errors, we use the spectroscopic sample and
set the redshift to the spectro-z value. Then, we determine the
best-fit template for each spectroscopic galaxy. For random, normally
distributed uncertainties in the flux measurements, $\overline{\Delta
F_f}$ should be $\sim$0 (where $\overline{\Delta F_f}$ is the average
difference between predicted and observed fluxes in the filter $f$).
Instead, we initially find systematic offsets of $\overline{\Delta
F_f}$ (as for earlier photo-z derivations, e.g. Brodwin et al. 2006;
Ilbert et al. 2006; Mobasher et al. 2007). Such offsets are mainly due
to:
\begin{itemize}
\item uncertainties in the absolute calibration of the photometric
 zero-points.
\item uncertainties in the color modeling (filter transmission
  curves, incomplete set of templates or an incorrect extinction curve). 
\end{itemize}

These systematic offsets were removed using the iterative procedure
detailed in Ilbert et al. (2006). For each filter, $f$, we estimate
the values $s_f$ which minimize $\overline{\Delta F_f}$.  After having
applied the corrections, $s_f$, in each band $f$, the systematic
offsets derived in a second iteration can change. The values converge
after 3 iterations and we adopted the systematic offsets listed in
Table.~\ref{shift}.

\subsection{Extinction law}

Ilbert et al. (2006) adopted the dust extinction law measured in the
Small Magellanic Cloud (Prevot et al. 1984). However, considerable
changes in the extinction curve are expected from galaxy to
galaxy. Maraston et al. (2006) considered as a free parameter the
different extinction curves of the ``Hyperz" package (Bolzonella et
al. 2000) (e.g. Milky Way, large and small Magellanic clouds and
Calzetti). In order to limit the risk of catastrophic failures, we
adopt an intermediate approach here, using the most suitable
extinction curve depending on the SED template.

For each galaxy of the zCOSMOS sample, we set the redshift to the
spectroscopic redshift value. Then, we determine the best fit-template
and the appropriate color excess $E(B-V)^{\rm best}$. In this fit, we
assume an extinction curve $k(\lambda)$ ($=A(\lambda)/ E(B-V)$). Since
the extinction curves do not differ strongly at $\lambda>3000\rm{\AA}$
(blue and green curves in Fig.~\ref{SEDext}), we fit the templates
using only passbands with $\lambda > 3000 (1+z) \rm{\AA}$. With this
procedure, the $E(B-V)^{\rm best}$ value does not depend significantly
on the adopted extinction curve.

The extinction curves differ strongly at
$\lambda<3000\rm{\AA}$. Therefore, we use the rest-frame observed SEDs
at $\lambda_{\rm rest-frame} < 3000 \rm{\AA}$ to discriminate between
the different extinction curves. Using $m^{\rm obs}$ to represent the
observed magnitude and $m^{template}_{uncor}$ to represent the
predicted magnitude from the best-fit template (uncorrected for
extinction), the extinction $A(\lambda)$ is given by
$A(\lambda)=m^{\rm obs} - m^{\rm template}_{\rm uncor}$.
Fig.~\ref{SEDext} shows the rest-frame $(m^{\rm obs} - m^{\rm
template}_{\rm uncor})/E(B-V)^{\rm best}$ for each flux measurement,
i.e the extinction curve $k(\lambda)$. These points are compared with
the Calzetti et al. (2000) and the Prevot et al. (1984) extinction
curves. (We scaled the Prevot et al. curve to the same $A_{\rm V}$
value as Calzetti et al. by applying a factor $4.05/2.72$ which is the
ratio between the $R_V$ of the Calzetti and Prevot laws.) The Small
Magellanic Cloud extinction curve (Prevot et al. 1994) is well suited
for galaxies redder than the starburst template SB3. For galaxies
bluer than SB3, the Calzetti et al. (2000) extinction curve is found
to be more appropriate (see Fig.~\ref{SEDext}). For the IR sources
which are strongly star-forming, Caputi et al. (2008) show also that
the Calzetti et al. (2000) extinction law is more appropriate. This
result is not surprising since the Calzetti law was determined from
observed starburst galaxies. A broad absorption excess at
2175$\rm{\AA}$ (UV bump) seems necessary to explain the UV flux in
some starburst galaxies. The presence of this UV bump can be seen in a
theoretical modeling of the Calzetti law (Fischera et al. 2004) and in
the K20 sample of high redshift galaxies at $1<z<2.5$ (Noll et
al. 2007).

To summarize, we apply an additional extinction to the templates
according to a grid $E(B-V)=0,0.05,0.1,0.15,0.2,0.25,0.3,0.4, 0.5$. We
use the Prevot et al. (1994) extinction curve for the templates redder
than SB3, and Calzetti et al. (2000) for the templates bluer than
SB3. We allow an additional bump at 2175$\rm{\AA}$ for the Calzetti
extinction law if it produces a smaller $\chi^2$.  No reddening is
allowed for galaxies redder than $Sb$.

\subsection{$\chi^2$ minimization}

The photo-z is the redshift value which minimizes the merit function
$\chi^2(z, T, A)$:
\begin{equation}
\chi^2=\sum_{f=1}^{N_f} \left( \frac{ F_{\rm obs}^f-A\times
F_{\rm pred}^f(z, T) \; 10^{-0.4 s_f}}{\sigma_{\rm obs}^f} \right)^2,
\label{chi2}
\end{equation}
where $F_{{\rm pred}}^f(T, z)$ is the flux predicted for a template
$T$ at redshift $z$. $F_{{\rm obs}}^f$ is the observed flux and
$\sigma_{\rm obs}^f$ is the associated error. The index $f$ refers to
each specific filter and $s_f$ is the zero-point offset listed in
Table.\ref{shift}.  The opacity of the inter-galactic medium (Madau et
al. 1995) is taken into account. The photo-z is estimated from the
minimization of $\chi^2$ with respect to the free parameters, $z$, $T$
and the normalization factor $A$.  The color excess $E(B-V)$ is
included in the term $T$ (see section 3.4). The grid spacing in
redshift is $\delta z=0.01$ and the final redshift is derived by
parabolic interpolation of the redshift probability distribution. The
redshift probability distribution function (PDFz) is derived directly
from the $\chi^2(z)$ distribution:
\begin{equation}
P(z) \propto exp(- \frac{\chi^2(z)-\chi^2_{\rm min}}{2}).
\end{equation}
The minimum and maximum redshifts around the photo-z solution,
corresponding to the 1$\sigma$ errors, are estimated from the equation
$\chi^2(z)=\chi^2_{\rm min}+1$. As in Ilbert et al. (2006), we
increased the SExtractor flux errors by a factor of 1.5. This factor
does not shift the best-fit photo-z value but broadens the $\chi^2$
peak and derived redshift uncertainty.

\begin{figure}[htb]
\includegraphics[width=8cm]{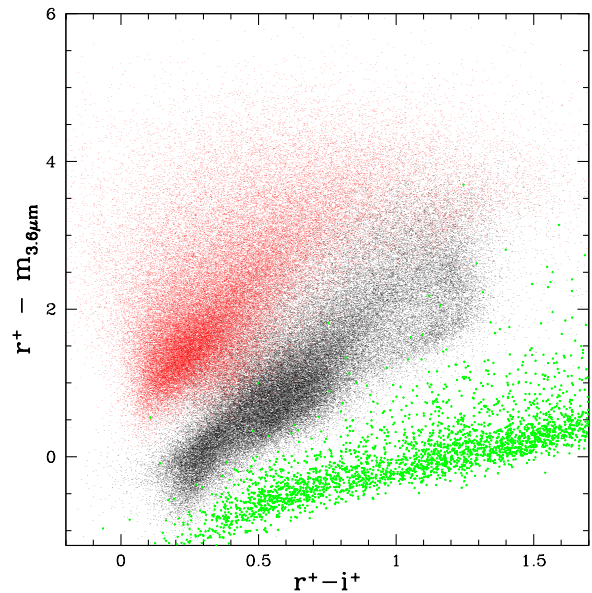}
\caption{Color-color diagnostics for sources including stars and galaxies. 
Stars revealed by the $\chi^2$ classification are shown in green; the
red and grey points are galaxies with $z_{\rm p}>1$ and $z_{\rm p}<1$,
respectively. \label{BzK}}
\end{figure}

\subsection{Star/AGN/galaxy classification}

For each object, $\chi^2$ is evaluated for both the galaxy templates
and stellar SED templates (Chabrier et al. 2000, Bixler et
al. 1991). If $\chi^2_{\rm gal}-\chi^2_{\rm star}>0$, where
$\chi^2_{\rm gal}$ and $\chi^2_{\rm star}$ are the minimum $\chi^2$
values obtained with the galaxy and stellar templates respectively,
the object is flagged as a possible star. Leauthaud et al. (2007)
catalogued point-like sources in the COSMOS field using the peak
surface brightness measured on the ACS images.  The SED $\chi^2$ and
morphological classification methods were found to be in excellent
agreement -- 84\% of the point-like sources at $i^+ <24$ are
classified as stars with the SED $\chi^2$ criterion mentioned above,
while only 0.2\% of the extended sources on the ACS images are
misclassified as stars with the same $\chi^2$ criterion. As shown in
Fig.~\ref{BzK}, the star sequence colors are well distinguished from
the bulk of the galaxy population if the colors include a NIR
band. NIR and MIR data are crucial to separate stars from
galaxies. Therefore, we limit the $\chi^2$ star classification to
$K<24$ or $F_{3.6 \mu m} > 1
\mu Jy$.  2\% of the point-like sources with $i^+ <24$ have $K>24$ 
and $F_{3.6 \mu m} > 1 \mu Jy$. These objects, which are only 2\% of
the total population at $i^+<24$, could be stars not recognized as
such by our SED $\chi^2$ classification.

AGN can be identified by their point-like X-ray emission. Most ($\sim
90\%$) of the 1887 sources detected in {\it XMM}-COSMOS (Brusa et
al. 2007) are dominated by an AGN.  Their photo-z determinations
require a different treatment: a correction for variability of the
photometric data and the use of SED templates specifically tuned to
AGN and their host galaxies. Accurate photo-z for the {\it XMM}-COSMOS
sources are derived in a companion paper (Salvato et al. 2008). Due to
the flux limit in XMM-COSMOS (Brusa et al. 2007), the moderately
luminous (log(Lx)=42-43 erg/s) and luminous (log(Lx)=43-44 erg/s) AGN
are not sampled by XMM-COSMOS at $z=0.5-1.25$ and $z=1.5-3$,
respectively. Therefore, this population is not identified as AGN in
the galaxy catalogue.

\section{Photometric redshift accuracy}

In this section, we assess the quality of the derived photometric
redshifts by two approaches: comparison with high confidence
spectroscopic redshifts and an analysis of the width of the redshift
probability distribution function obtained from the $\chi^2$ fitting
for the photo-z. The latter approach is justified by the excellent
agreement of the z$_s$-z$_p$ distribution with the width of the
probability distribution function for the spectroscopic sample.

\begin{figure*}
\begin{tabular}{c c}
\includegraphics[width=8cm]{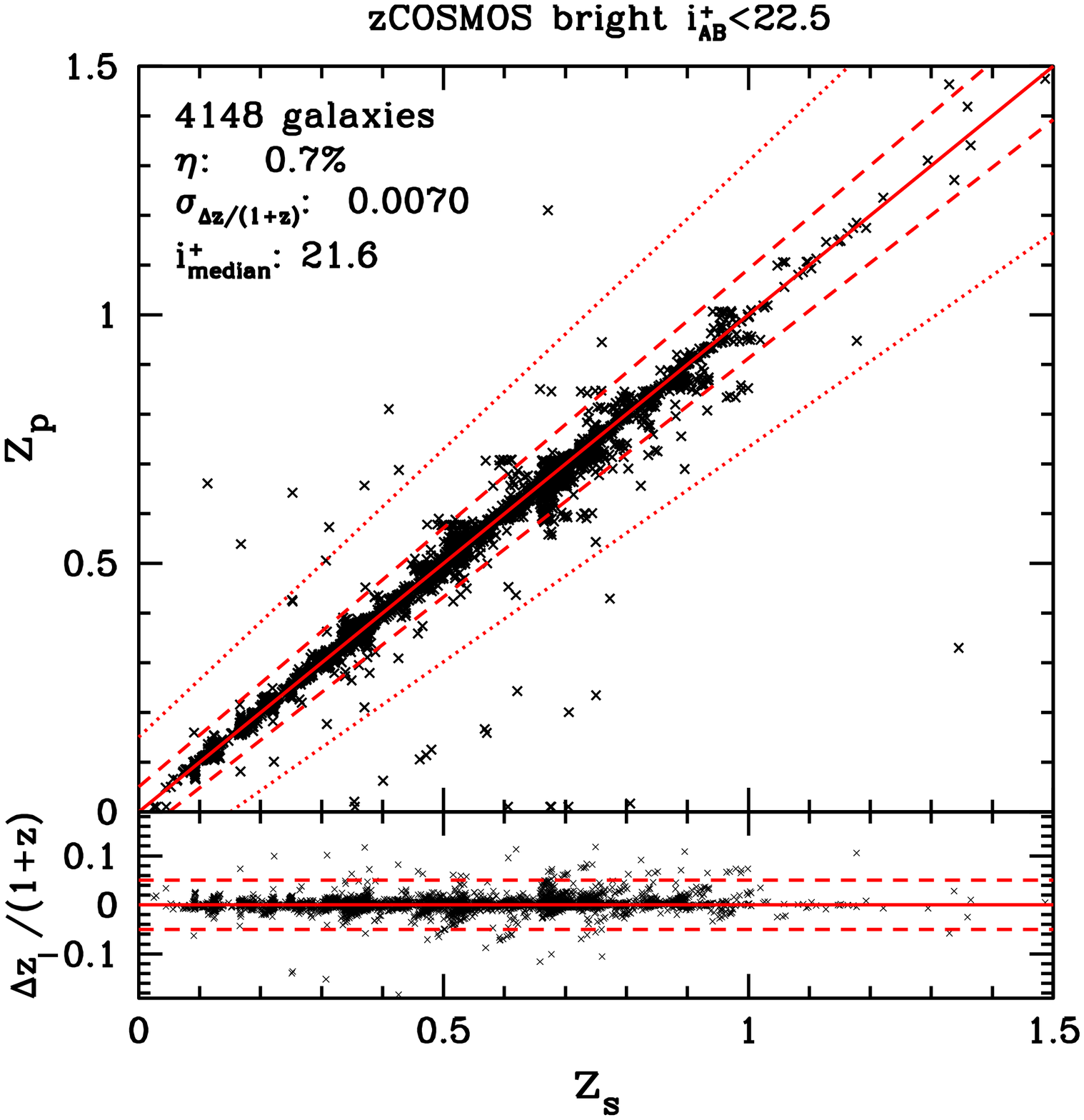}  &
\includegraphics[width=8cm]{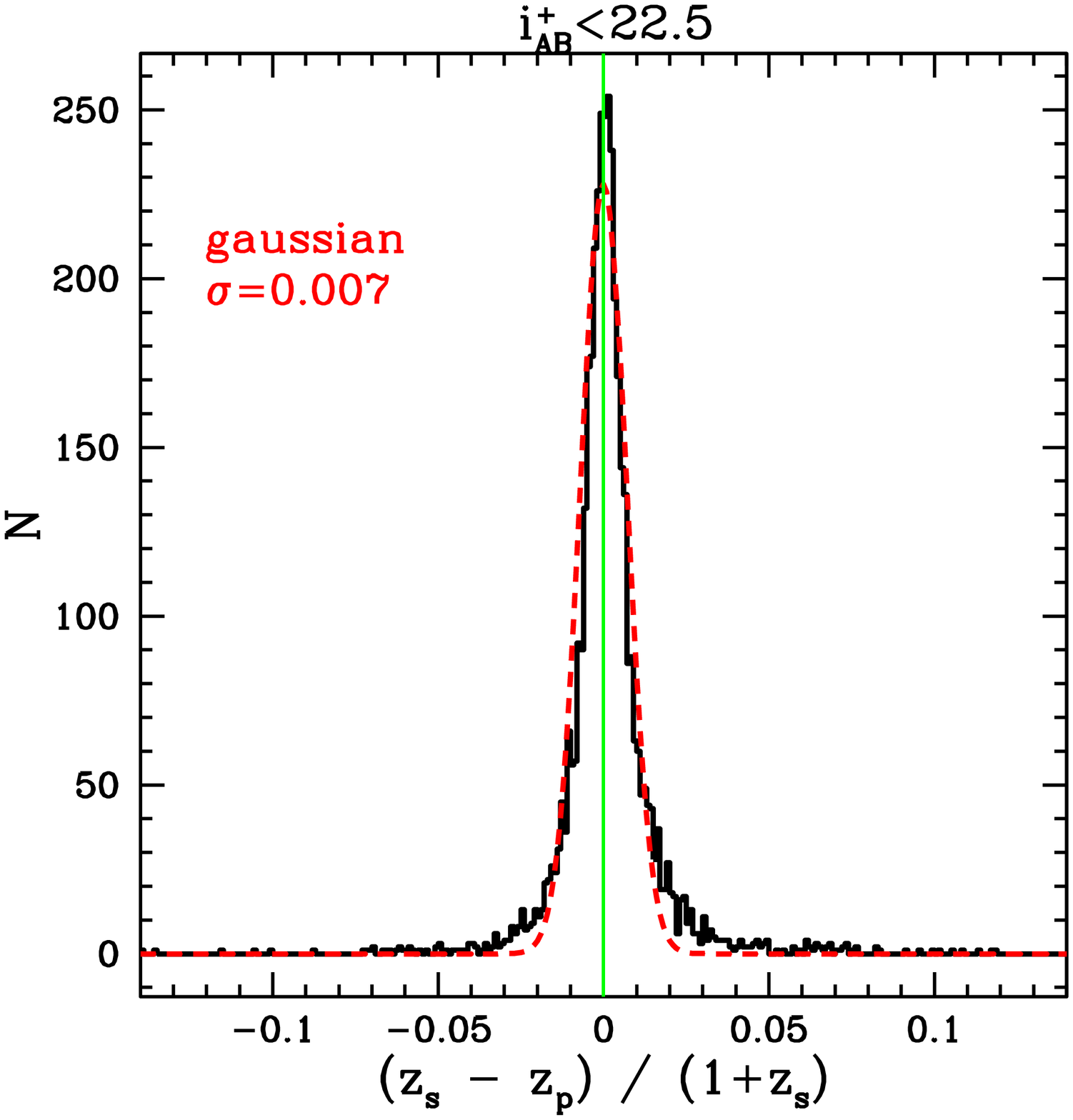} \\
\end{tabular}
\caption{Left panel: Comparison between $z_{\rm p}$ and $z_{\rm s}$ 
 for the bright spectroscopic selected sample $17.5 \le i^+_{\rm AB}
 \le 22.5$ (zCOSMOS-bright: Lilly et al. 2008). The dotted and dashed
 lines are for $z_{\rm p}=z_{\rm s} \pm 0.15 (1 + z_{\rm s})$ and
 $z_{\rm p}=z_{\rm s} \pm 0.05 (1 + z_{\rm s})$, respectively. The
 1$\sigma$ dispersion, the fraction of catastrophic failures and the
 median apparent magnitude are listed in the top left corner of the
 left panel. Right panel: $\Delta z/(1+z_{\rm s})$ distribution. The
 dashed line is a gaussian distribution with $\sigma=0.007$.
\label{zp_zs_bright}}
\end{figure*}

\begin{figure*}
\begin{tabular}{c c}
\includegraphics[width=8cm]{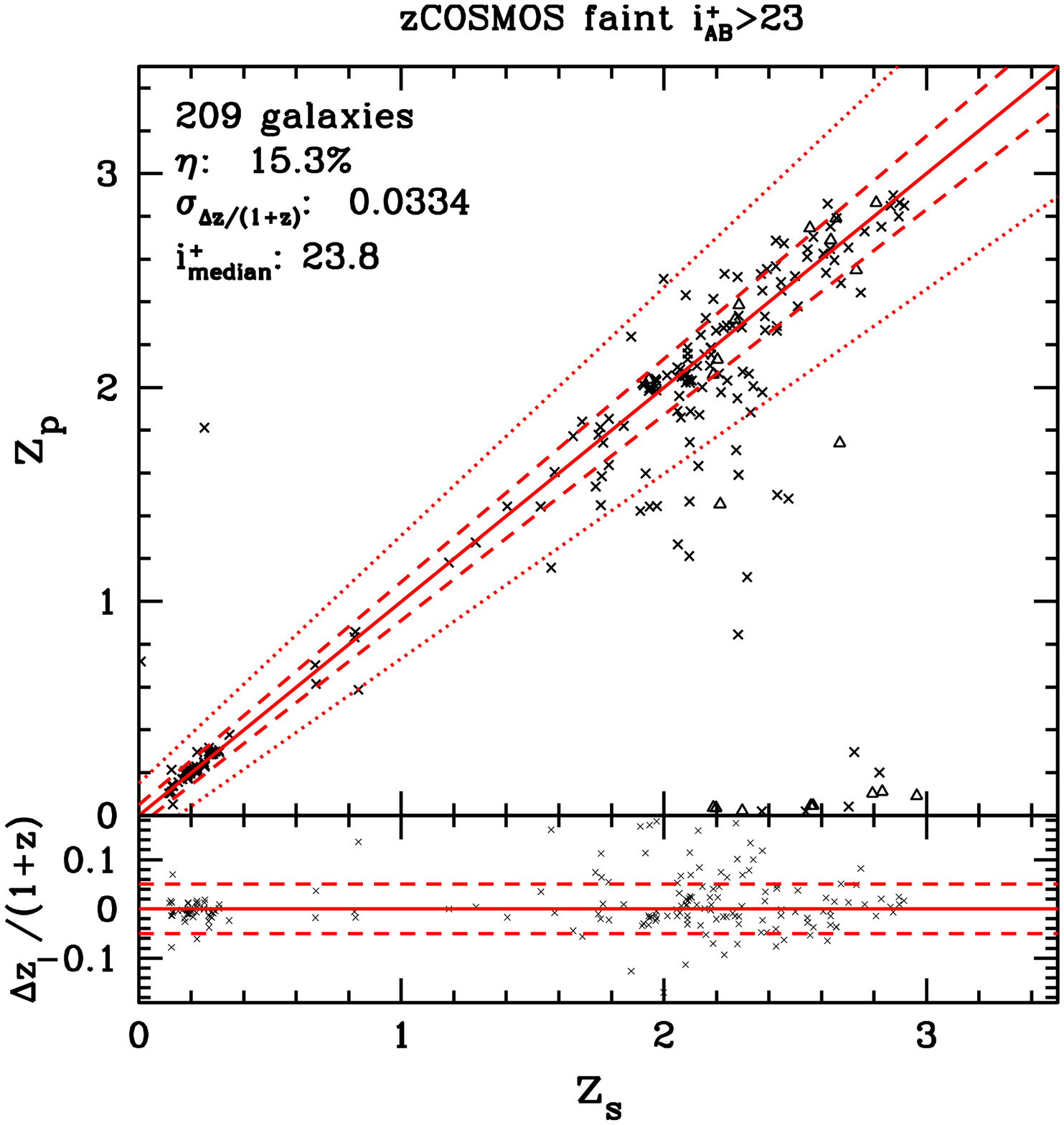}  &
\includegraphics[width=8cm]{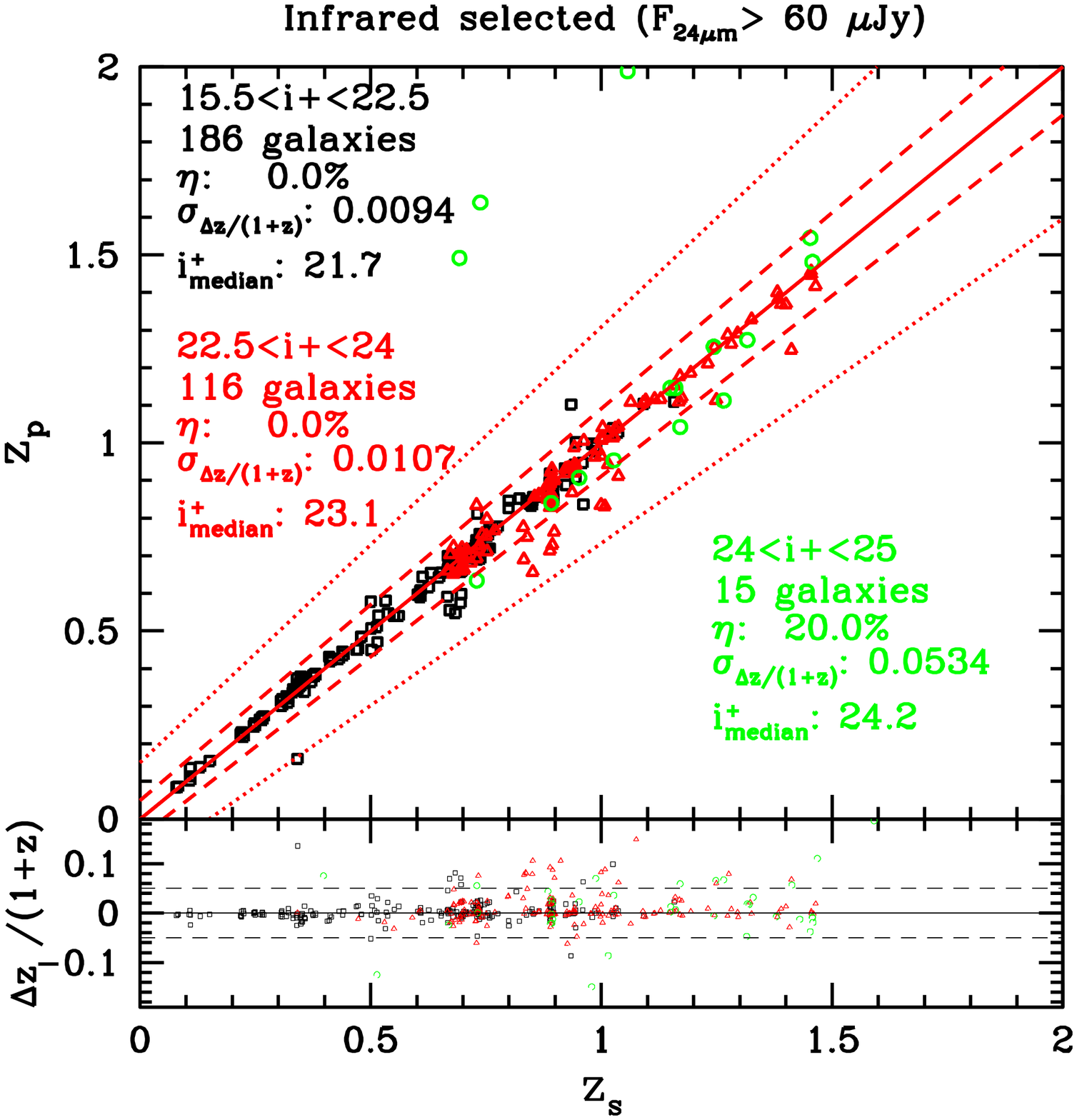} 
\end{tabular}
\caption{Comparison between $z_{\rm p}$ and $z_{\rm s}$. 
 \ \ Left panel: zCOSMOS-faint sample (Lilly et al. 2008). The open
 triangles are objects with a secondary peak in the redshift
 probability distribution function. \ \ Right panel: infrared selected
 sample (Kartaltepe et al. 2008) split into a bright sample $i^+ <
 22.5$ (black), a faint sample $22.5<i^+ <24$ (red) and a very faint
 sample $24<i^+ <25$ (green).
\label{zp_zs}}
\end{figure*}

\subsection{Comparison of Photometric and Spectroscopic Redshifts}

We first assess the quality of the photo-z by comparison with the
spectro-z. If $\Delta z = z_{\rm s} - z_{\rm p}$, we can estimate the
redshift accuracy from $\sigma_{\Delta z /(1+z_{\rm s})}$ using the
normalized median absolute deviation (NMAD: Hoaglin et al. 1983)
defined as $1.48 \; \times \; {\rm median}(| z_{\rm p} -z_{\rm s}
|/(1+z_{\rm s}))$. The NMAD is directly comparable to other papers
which directly quote the $rms/(1+z)$. This dispersion estimate is
robust with respect to catastrophic errors (i.e., objects with $|
z_{\rm p}-z_{\rm s} |/(1+z_{\rm s}) > 0.15$). The percentage of
catastrophic errors is denoted by $\eta$.

Figure~\ref{zp_zs_bright} (left panel) shows the comparison between
$z_{\rm p}$ and $z_{\rm s}$ for the zCOSMOS-bright sample selected at
$i^+<22.5$. The spectro-z sample is selected only by apparent
magnitude and is therefore representative of the entire $i^+<22.5$
population. We obtain an accuracy of $\sigma_{\Delta z/(1+z)}=0.007$
at $i^+<22.5$ and the distribution of offsets ($(z_{\rm p}-z_{\rm
s})/(1+z_{\rm s})$) is well fit by a gaussian with $\sigma=0.007$
(Fig.~\ref{zp_zs_bright} right panel). The percentage of catastrophic
failures is below 1\%.

The zCOSMOS-faint sample (Fig.~\ref{zp_zs}, left panel) with a median
apparent magnitude of $i_{\rm med}^+\sim 24$ provides a quality check
for the photo-z at $1.5<z<3$ where the photo-z are expected to have a
significantly higher uncertainty.  The faint sample includes galaxies
at $i^+\sim 25$ and the Balmer break is shifted into the NIR where the
filter set has gaps. At $1.5<z<3$, the accuracy is found to be
$\sigma_{\Delta z/(1+z)}=0.06$ with 20\% catastrophic failures. The
fraction of high-redshift galaxies ($z_{\rm s}>1.5$) for which a
low-redshift photo-z ($z_{\rm p}<0.5$) was assigned is $7\%$ but this
failure rate drops to $4\%$ if the sample is restricted to galaxies
detected at $IRAC(3.6\mu m) >1 \mu Jy$. The zCOSMOS-faint sample
actually includes 54 galaxies with low spectroscopic redshift ($z_{\rm
s}<0.5$ ); 53 out of the 54 galaxies ($>98$\%) were assigned the
correct photo-z low redshift. In summary, we conclude that, for the
faint galaxies, the photo-z of low redshift objects are still assigned
correctly, while the failure rate for high redshift objects is
significantly reduced if the objects have good IRAC detections. Such
result is expected since the photo-z are already including the same
information present in high-z color selections such as $BzK$ (Daddi et
al. 2004) or similar diagnostics using the IRAC bands (Fig.~\ref{BzK})
to isolate the good redshift range.

The zCOSMOS-bright and zCOSMOS-faint samples do not probe $0.2<z<1.5$
at $i^+>22.5$. Here, we use as a comparison sample the MIPS-spectro-z
(Kartaltepe et al. 2008).  In Fig.~\ref{zp_zs} (right panel), we split
the MIPS-spectro-z sample into bright ($i^+ <22.5$), faint ($22.5<i^+
<24$) and very faint samples ($24<i^+ <25$).  For the bright
sub-sample, the dispersion of the MIPS-selected galaxies is
$\sigma_{\Delta z/(1+z_{\rm s})}=0.009$, only slightly greater than
that of the optically selected sample at $i^+ <22.5$.  For the faint
sub-sample (median apparent magnitude $i^+
\sim 23.1$), $\sigma_{\Delta z /(1+z_{\rm s})}=0.011$, i.e. slightly
worse than for the brighter optically-selected objects.  For the very
faint sub-sample, the accuracy is degraded to $\sigma_{\Delta z
/(1+z_{\rm s})}=0.053$ with a catastrophic failures rate of 20\%.
This degradation is due to decreasing signal-to-noise photometry for
faint objects and could be amplified by the infrared selection which
picks up more heavily obscured galaxies and higher redshift galaxies
(e.g. Fig.4 of Le Floc'h et al. 2005).

\begin{figure}[ht]
\includegraphics[width=8cm]{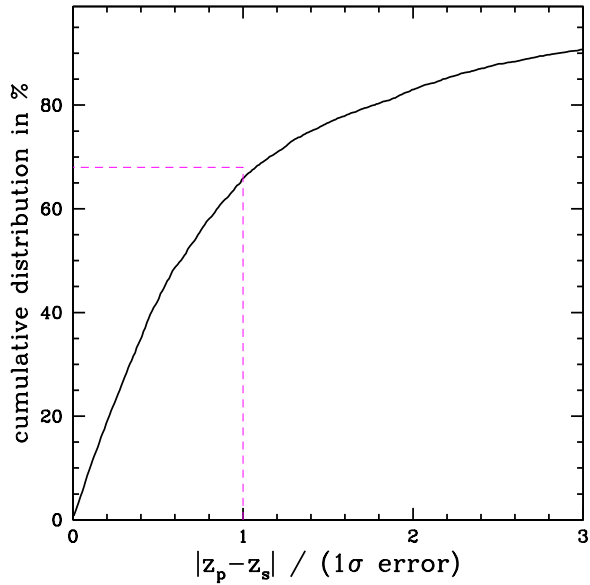}
\caption{Cumulative distribution of the ratio
$|z_{\rm p}-z_{\rm s}|/ (1\sigma \; {\rm error})$. 65\% of the photo-z
have a spectro-z solution encompassed within the 1$\sigma$ error,
close to the expected value of 68\% (magenta dashed line).
\label{error68}}
\end{figure}

\begin{figure}[ht]
\includegraphics[width=8cm]{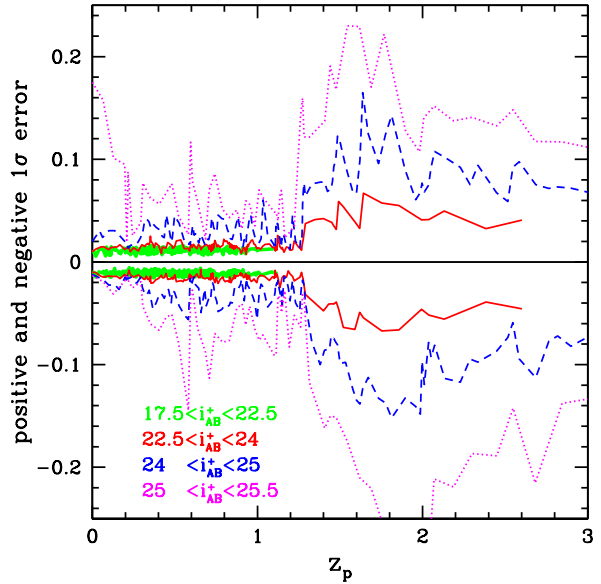}
\caption{1$\sigma$ uncertainty for  the $z_{\rm p}$ estimate as a function of
  redshift in different apparent magnitude bin. Each value is computed
  with 50 galaxies per bin.  The thick solid green lines, the solid
  red line, the dashed blue lines and the dotted magenta lines are for
  $i^+<22.5$, $22.5<i^+<24$, $24<i^+<25$, $25<i^+<25.5$, respectively.
\label{error_zp}}
\end{figure}

\subsection{Accuracy derived from the Photo-z Probability Distribution Function}

Since evaluation of the photo-z accuracy from the comparison with
spec-z is limited to specific ranges of magnitude and redshift, we use
the 1$\sigma$ uncertainty in the derived photo-z probability
distribution to extend the uncertainty estimates over the full
magnitude/redshift space.

The reliability of the 1$\sigma$ uncertainty estimate for the photo-z
as derived from the Probability Distribution Function (PDFz) (see
\S3.5) can be checked by comparing this uncertainty with that
derived directly from the photo-z - spec-z offsets for the
spectroscopic sample. Figure~\ref{error68} shows the cumulative
distribution of these offsets normalized by the 1$\sigma$ uncertainty
in the probability function for the zCOSMOS-bright sample (the ratio
$|z_p-z_s|/(1\sigma \; error)$ is lower than 1 if the measured offset
$z_p-z_s$ is lower than the 1$\sigma$ uncertainty). 65\% of the
$z_{\rm p}$ are within the 1$\sigma$ error bars, whereas the expected
fraction is 68\%. We therefore conclude that the 1$\sigma$
uncertainties in the probability function as derived here provide a
robust assessment of the accuracy in $z_{\rm p}$.

Figure~\ref{error_zp} shows the 1$\sigma$ negative and positive
uncertainties derived from the probability function as a function of
redshift and apparent magnitude. Two clear conclusions emerge: the
accuracy is inevitably degraded for fainter galaxies at all redshifts
and the photo-z have significantly higher uncertainty at $z\gtrsim
1.25$.

\begin{figure*}[htb]
\begin{tabular}{c c}
\includegraphics[width=8cm]{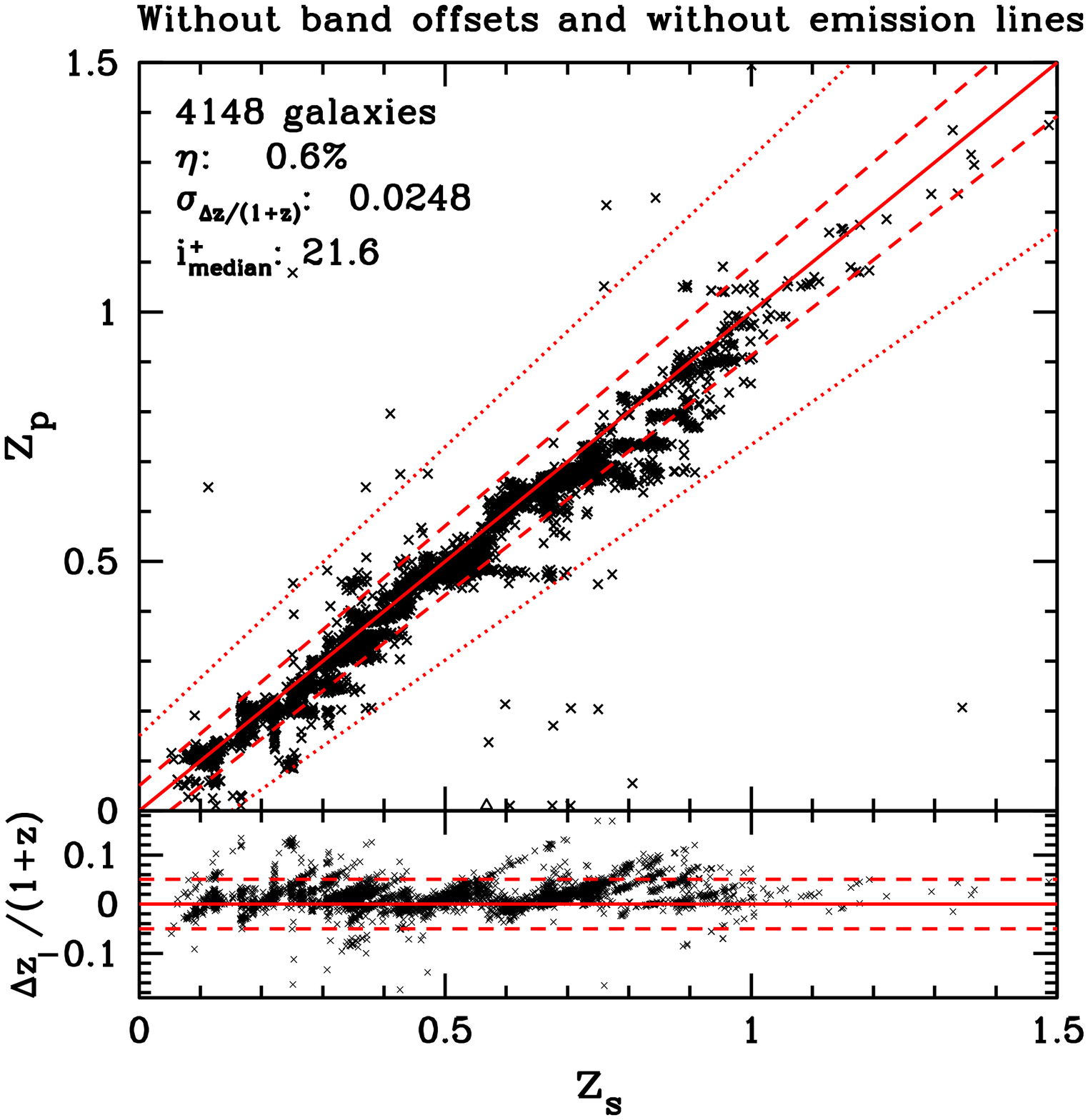} &
\includegraphics[width=8cm]{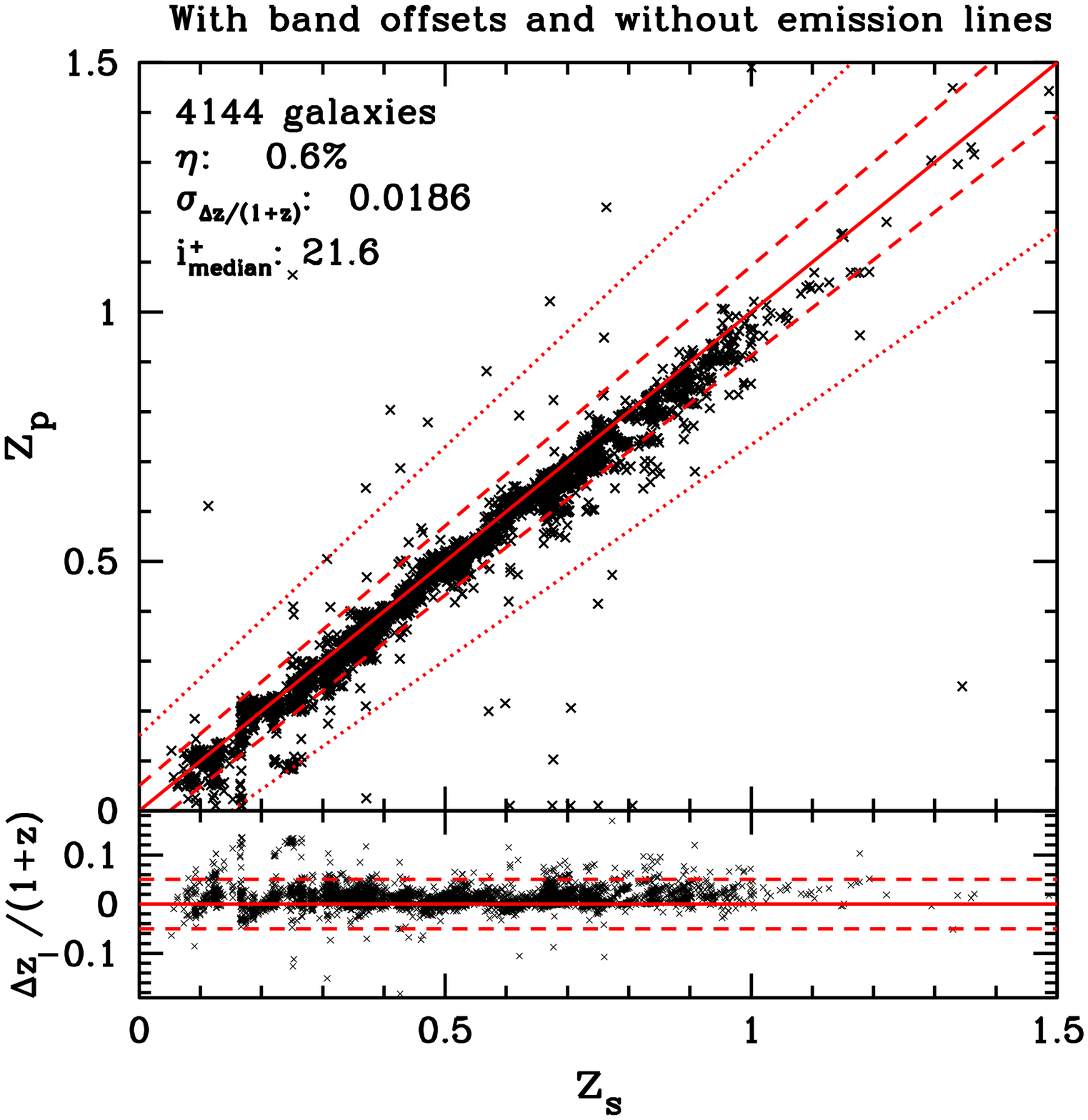} 
\end{tabular}
\caption{Comparison between $z_{\rm p}$ and $z_{\rm s}$ at different steps of 
the method. The comparison is done for the zCOSMOS-bright sample
selected at $17.5 \le i^+ \le 22.5$. The left panel shows the photo-z
computed with a standard $\chi^2$ method (no emission lines
contributions and without calibration of the band offsets). The right
panel shows $z_{\rm p}$ computed with the calibration of the band
offsets, but without including emission lines in the
templates. Including emission lines improves the accuracy by a factor
$\sim$2.5. \label{zp_zs_method}}
\end{figure*}

From $z>0$ out to $z=1.25$, the 1$\sigma$ errors do not depend
significantly on the redshift, with $\sigma_{\Delta z}
\lesssim 0.02$ at $i^+ <24$ (note that we dropped out the division of $\sigma$ 
by $(1+z)$ in this analysis). To first order, the photo-z are accurate
when the wavelengths of the Balmer and/or Lyman breaks are well
constrained. Therefore, the wavelength coverage of the filter set and
the photometry sensitivities determine the photo-z accuracy as a
function of redshift. The COSMOS photometry coverage is continuous and
dense in the optical from the $u^*$ band ($\lambda_{\rm eff}\sim
3911{\rm \AA}$) to the $z^+$ band ($\lambda_{\rm eff}\sim 9021{\rm
\AA}$). The average wavelength spacing between consecutive filters is
also only 230\AA. The Balmer break at redshift $z=0$ out to $z=1.25$
is always at $\lambda<9000{\rm \AA}$, thus the high accuracy in this
redshift range.

The accuracy degrades at $z>1.25$, where the 4000${\rm \AA}$ Balmer
break goes out of the $z^+$ band ; at $z\sim 1.8$ it is a factor 3
larger than at $z\sim 1.25$ ($\sigma_{\Delta z} \sim 0.14$ for $i^+
\sim 24$). The lack of coverage between the $z'$ and the $J$ band
accounts for the discontinuous increase in uncertainty at $z>1.25$
since the Balmer break can not be located with precision. At $z>1.5$,
the estimated error from the photo-z -- spectro-z comparison is
$\sigma_{\Delta z}=0.19$ (see \S4.1; $\sigma_{\Delta z/(1+z_{\rm
s})}=0.06$ at the median redshift $z\sim 2.2$). This accuracy is $\sim
1.4$ bigger than that estimated from the 1$\sigma$ uncertainty
($\sigma_{\Delta z}\sim 0.14$ at $i^+\sim 24$). Errors due to bias in
photo-z will not be recovered from the probability distribution
function which could explain this difference. At $z>1.25$, all the
optical filters, which are about 80\% of all used filters, are
sampling the rest-frame UV at $\lambda<3500 {\rm \AA}$. Uncertainties
in the adopted extinction law have considerable impact on the UV slope
and could introduce small biases in the estimate of the photo-z.

At $z>2.5$, the accuracy improves improves again ($\sigma_{\Delta z}
\lesssim 0.1$ for $24<i^+<25$) when the 4000${\rm \AA}$ Balmer break enters 
in the $J$ band ($z\sim 2$) and the UV light shortward of $L_{\alpha}$
enters the $u^*$ band ($z\sim 2.3$).

\subsection{Importance of zero-point offsets and emission lines}

The major improvements in the technique implemented here are the
carefully iterative evaluation of the photometric zero-point offsets
for all bands and the allowance for a range of emission line
contributions to the fluxes of template SEDs.

The left panel of Fig.~\ref{zp_zs_method} shows the comparison between
$z_{\rm p}$ and $z{\rm s}$ computed without correction of the
systematic band offsets (see \S3.3) and without including the emission
lines (see \S3.2). Some systematic biases (horizontal and vertical
stripes in Fig.~\ref{zp_zs_method}) greater than $\delta_{z}\sim 0.1$
are clearly seen in the photo-z estimate. For example, galaxies with
$z_{\rm s} \sim 0.8$ are often shifted to $z_z \sim 0.7$.  With the
iterative calibration of the band offsets turned on (right panel of
Fig.~\ref{zp_zs_method}), these biases are limited to
$\delta_z<0.05$. This clearly demonstrates the importance of
zero-point calibration to reduce the photo-z bias (as already shown by
Brodwin et al. 2006 and Ilbert et al. 2006).

The right panel of Fig.~\ref{zp_zs_method} shows the comparison
between $z_{\rm p}$ and $z_{\rm s}$ without including the emission
lines (see \S 3.2). The accuracy is $\sigma_{\Delta z/(1+z_{\rm s})}
\sim 0.02$. When the emission lines are included in the templates, the
accuracy is improved by a factor of 2.5 ($\sigma_{\Delta z/(1+z_{\rm
s})}=0.007$, left panel of Fig.~\ref{zp_zs_bright}). Therefore,
including emission lines in the SEDs is crucial, especially when the
medium bands are used to measure $z_{\rm p}$.

\begin{figure}[htb]
\includegraphics[width=8cm]{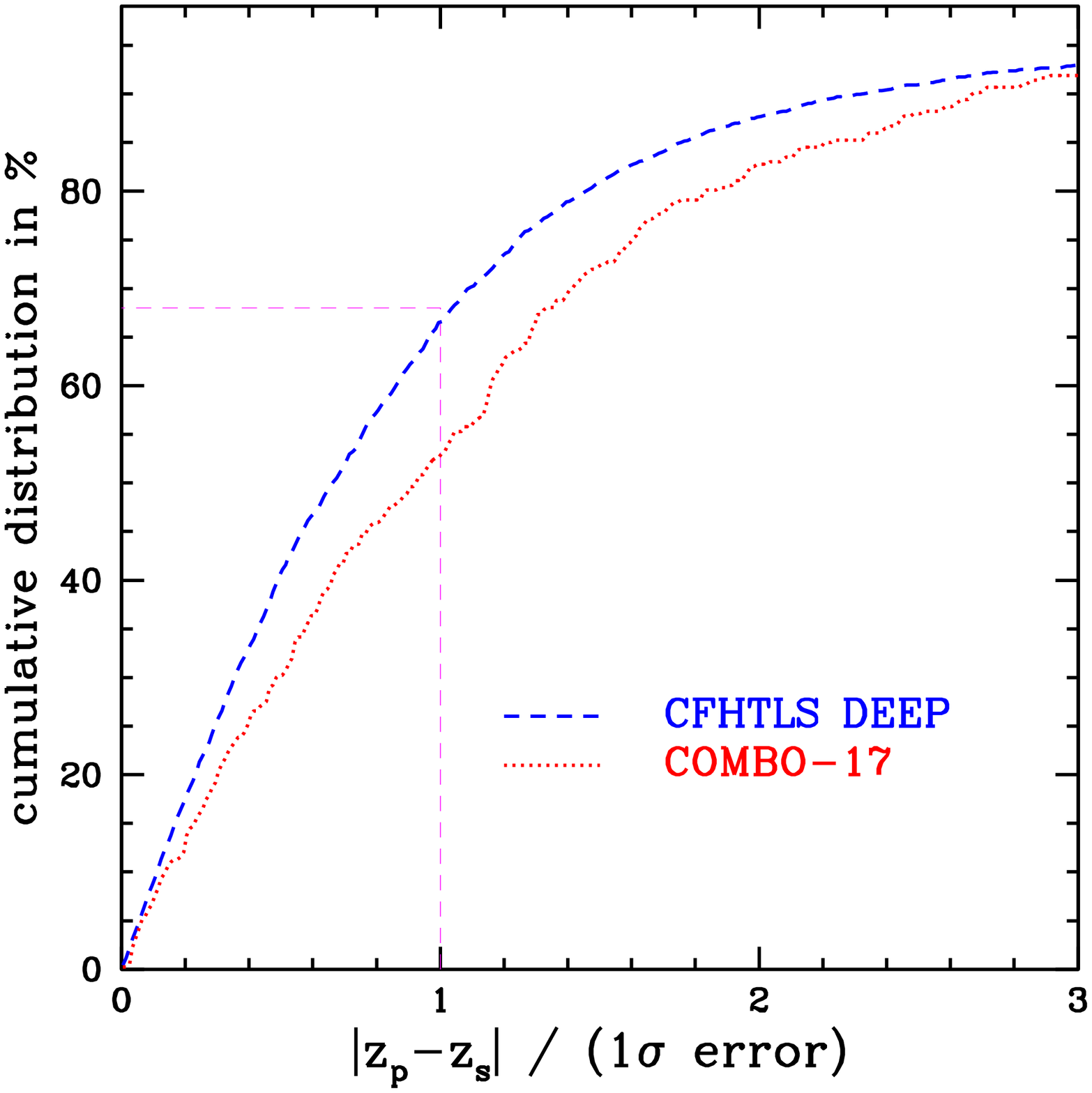}
\caption{Same as Fig.\ref{error68} for the
  CFHTLS-Deep survey (Ilbert et al. 2006) and the COMBO-17 survey
  (Wolf et al. 2004).
\label{cumule68}}
\end{figure}

\begin{figure}[htb]
\centering
\includegraphics[width=8.cm]{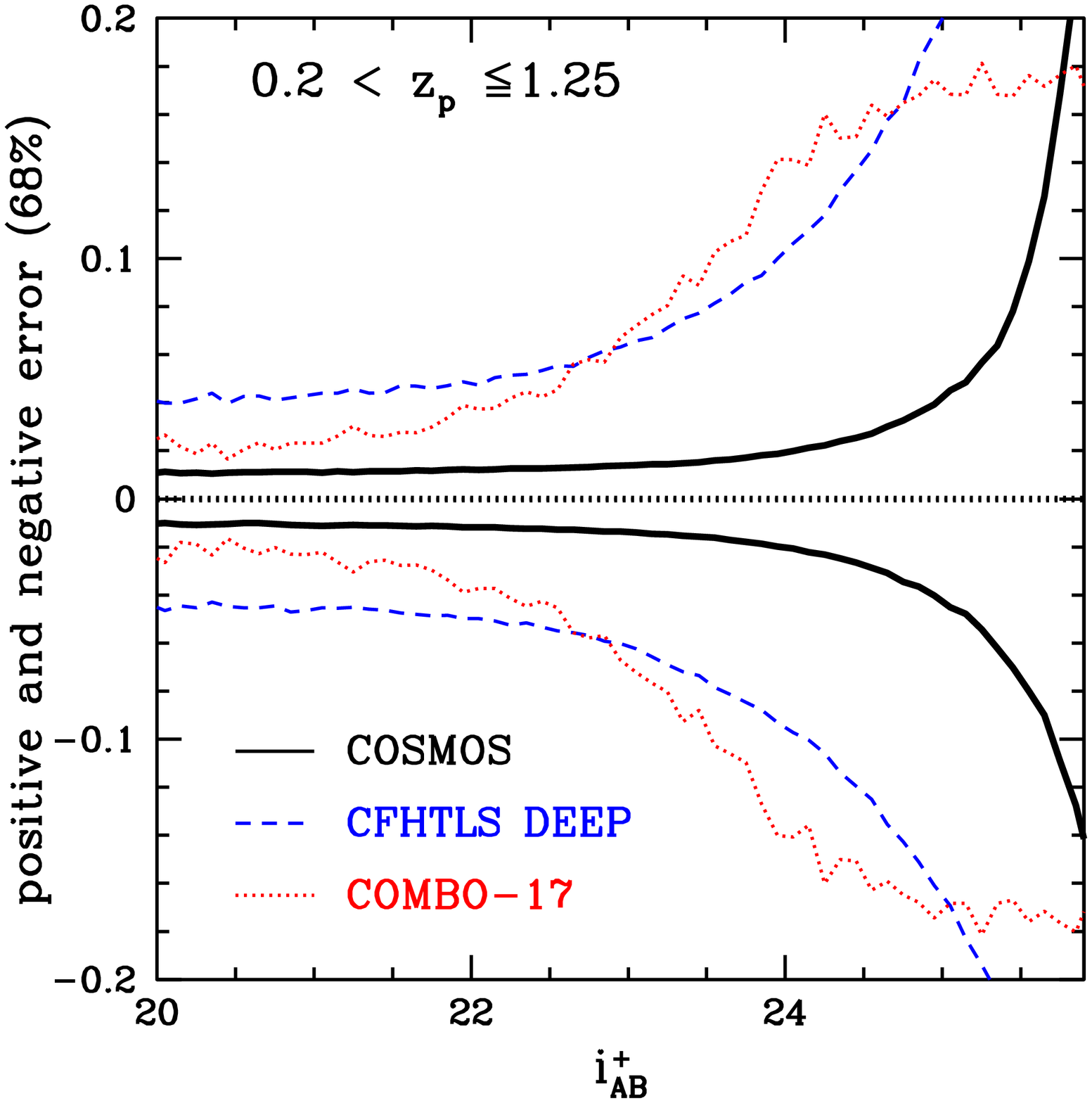}
\caption{1$\sigma$ error for the $z_{\rm p}$ estimate as a function of the
  apparent magnitude in the redshift range $0.2<z<1.25$. The 1$\sigma$
  errors have been rescaled by a factor 1.2 for the COMBO-17 survey
  (Wolf et al. 2004) (see text and Fig.~\ref{cumule68}).
\label{error_Ip}}
\end{figure}

\section{Discussion}

\subsection{Comparison with Other Surveys}\label{other-surveys}

We now compare the accuracy of the COSMOS-30 photo-z (this work) with
those obtained previously for COMBO-17 (Wolf et al. 2004) and
CFHTLS-DEEP (Ilbert et al. 2006).  The accuracies can be compared as a
function of apparent magnitude using the 1$\sigma$ measured
uncertainties in the photo-z probabilities. Once again, we check also
the validity of the 1$\sigma$ uncertainties for other surveys
following the method described in \S4.2.  For the COMBO-17 and CFHTLS
photo-z, we use the spectro-z from the VIMOS-VLT Deep Survey (Le
F\`evre et al. 2004, 2005).  Figure~\ref{cumule68} shows that about
53\%, 67\% of the values for $z_{\rm s}$ are included inside the
1$\sigma$ uncertainties for COMBO-17 and CFHTLS-DEEP survey,
respectively. The 1$\sigma$ errors for $z_{\rm p}$ in COMBO-17 are
rescaled by $\sim$1.2 to obtain 68\% of the $z_s$ within the $1\sigma$
error (This rescale is small and changes nothing in our conclusions).

Figure~\ref{error_Ip} shows the redshift dependence of the 1$\sigma$
uncertainties in photo-z as a function of magnitude for
$0.2<z<1.25$. The $z_{\rm p}$ for the CFHTLS-DEEP were derived from 5
broad bands ($u^*$, $g^\prime$, $r^\prime$ , $i^\prime$ ,
$z^\prime$). The accuracy of the COSMOS photo-z is improved by a
factor of 3 when compared to CFHTLS-DEEP. This improvement is largely
due to the 12 medium bands in this redshift range (NIR data have a
real impact only at $z>1.25$). We checked this by deriving photo-z
without the 12 medium bands, obtaining an accuracy of $\sigma_{\Delta
z/(1+z_{\rm s})}=0.03$ at $i^+ <22.5$ (similar to the COSMOS release
of Mobasher et al. 2007 using 8 broad bands $u^*$, $B_{\rm J}$,
$V_{\rm J}$, $g^+$, $r^+$, $i^+$, $z^+$ and $K_S$). COMBO-17 includes
12 medium bands in addition to the 5 broad bands and Wolf et
al. (2004) achieved an accuracy of $\sigma_{\Delta z/(1+z)}=0.02$ at
$i^+ \sim 21.5$. The accuracies of COMBO-17 photo-z are intermediate
between those of COSMOS-30 and CFHTLS for $i^+_{AB}<22.5$ but become
larger that the CFHTLS accuracies at fainter magnitudes, because the
COMBO-17 data are about 1.5{\ts}mag shallower than the CFHTLS data.
Since only secure photo-z (one solution) have been kept in the
COMBO-17 catalogue, it explains the flattening of the error bars at
$i^+_{AB}>24$.

\begin{figure}[htb]
\centering
\includegraphics[width=8.cm]{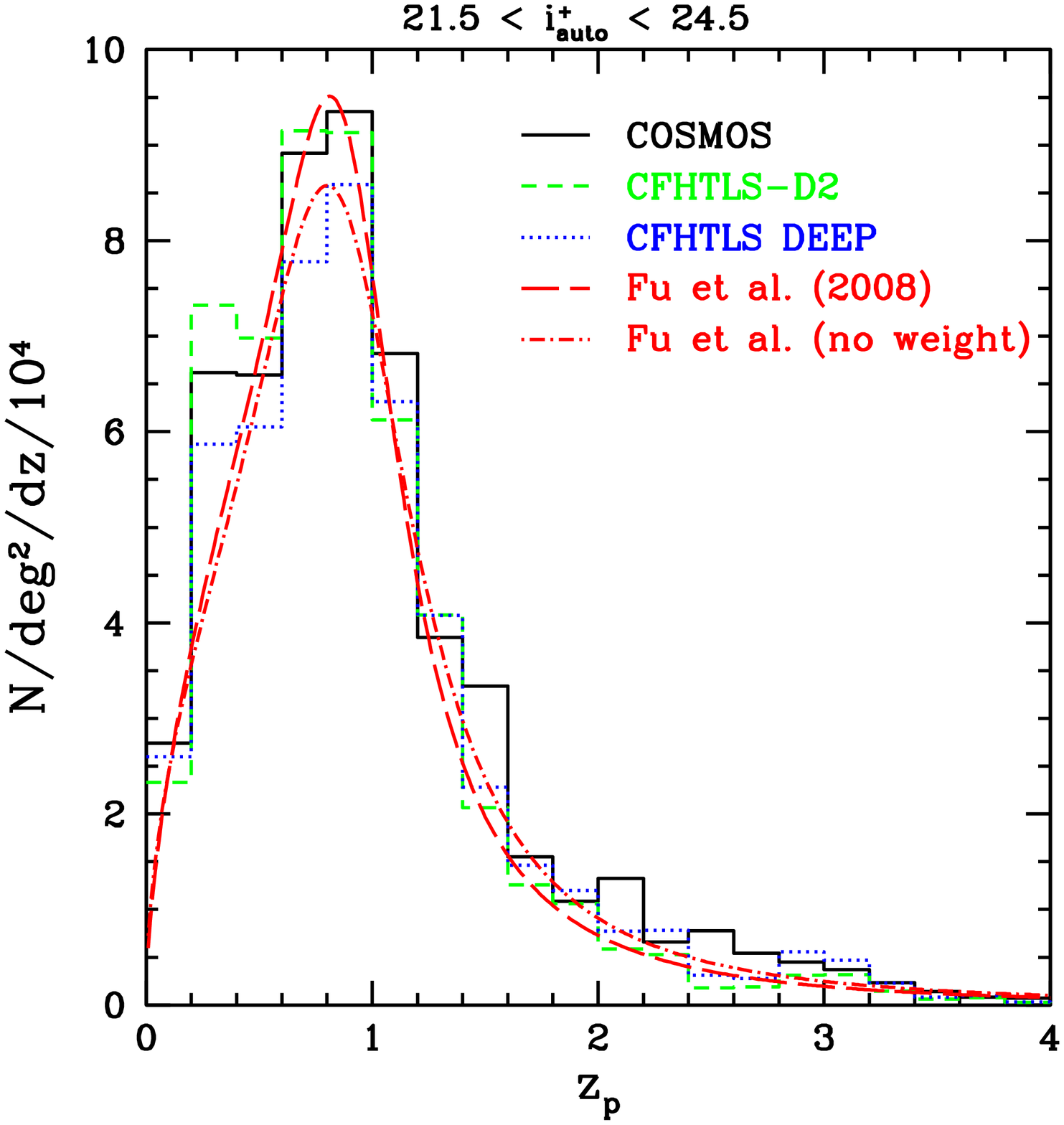}
\caption{Redshift distribution from the 2-deg$^2$ COSMOS survey (black solid
  line), from the 1-deg$^2$ CFHTLS-D2 field (green dashed line), from
  the 4-deg$^2$ CFHTLS-DEEP fields (blue dotted line, including also
  D2). CFHTLS-D2 covers 1 deg$^2$ within the COSMOS field. The red
  long dashed line is the redshift distribution obtained by Fu et
  al. (2008) who fit the CFHTLS-DEEP photo-z in the magnitude bin
  $21.5<i^+ <24.5$ and the dashed-dotted line is obtained without the
  weight applied for the CFHTLS weak lensing selection (J. Coupon,
  private communication). \label{distz_convoluee}}
\end{figure}

\begin{figure}[htb]
\centering
\includegraphics[width=8.cm]{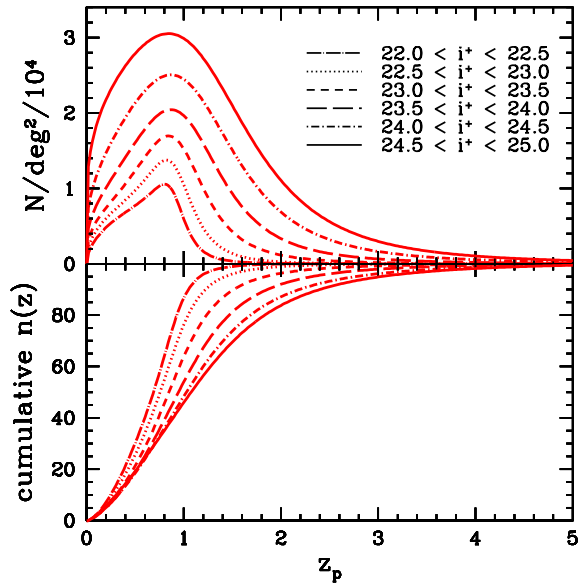}
\caption{Top panel: Evolution of the redshift distribution as a function of $i^+$ magnitude in the COSMOS field. We use the parametrization of Fu et al. (2008) in different redshift bins. Bottom panel: Cumulative redshift distribution. \label{distz_cosmos}}
\end{figure}

\begin{table*}
\begin{tabular}{ c c c c c c c c } \hline \\
             mag range        &    $a$       &        $b$     &     $c$  & $A$ & average z & median z\\
\\
\hline\\
 $22.0<i^+_{auto}<22.5$  &    0.497$\pm$0.019  & 12.643$\pm$0.409  &  0.381$\pm$0.016  &     4068.19   &   0.66 &  0.66  \\ 
 $22.5<i^+_{auto}<23.0$  &    0.448$\pm$0.016  &  9.251$\pm$0.218  &  0.742$\pm$0.030  &     9151.98   &   0.76 &  0.72  \\
 $23.0<i^+_{auto}<23.5$  &    0.372$\pm$0.012  &  6.736$\pm$0.094  &  1.392$\pm$0.055  &    18232.24   &   0.90 &  0.82  \\
 $23.5<i^+_{auto}<24.0$  &    0.273$\pm$0.008  &  5.281$\pm$0.039  &  2.614$\pm$0.096  &    35508.58   &   1.05 &  0.92  \\
 $24.0<i^+_{auto}<24.5$  &    0.201$\pm$0.005  &  4.494$\pm$0.024  &  3.932$\pm$0.134  &    60306.30   &   1.18 &  1.00  \\
 $24.5<i^+_{auto}<25.0$  &    0.126$\pm$0.003  &  4.146$\pm$0.021  &  5.925$\pm$0.191  &   103340.04   &   1.25 &  1.06  \\
\hline
\end{tabular}
\caption{Galaxy redshift distribution per deg$^2$. The parameters $a$, $b$, $c$ and $A$ of the eqn.\ref{NZ} function (Fu et al. 2008) are given per apparent bin.}
\label{fu}
\end{table*}

\subsection{Redshift Distribution of Galaxies}

Fig.~\ref{distz_convoluee} shows the galaxy redshift distribution per
deg$^2$ from the COSMOS and CFHTLS surveys (for $21.5<i^+_{auto}
<24.5$).  Although the surveys are largely independent and the photo-z
are computed with different codes, the overall agreement in the
redshift distributions is excellent. The agreement is particularly
good between COSMOS and CFHTLS-D2 which covers 1 deg$^2$ within the
COSMOS field.

The density of galaxies at $z>1.5$ in the CFHTLS fields is of a
crucial interest for weak lensing analysis (Benjamin et al. 2007). The
CFHTLS redshift distribution presents a bump at $z\sim 3$, but this
excess could be due to misidentifications between the $z<0.4$ and
$z>1.5$ photo-z (Van Waerbeke et al. 2008) when no NIR data are
available (as is the case for CFHTLS). The COSMOS photo-z are computed
with NIR data and such catastrophic failures are limited (see \S4.1).
Fig.~\ref{distz_convoluee} shows that this bump seems created by a
deficit of galaxies at $2.5<z<3$, by comparing CFHTLS-D2 and COSMOS.

Fu et al. (2008) used the CFHTLS photo-z distribution to estimate the
matter density parameter $\Omega_{\rm m}$ and the amplitude of the
matter power spectrum $\sigma_8$.  The redshift distribution from Fu
et al. (2008) at $21.5<i^+ <24.5$ and $z<2.5$ is over-plotted in
Fig.~\ref{distz_convoluee} with and without (J. Coupon, private
communication) the weight applied for the weak lensing selection
(dashed and dashed-dot lines, respectively). Given the 20\% variation
expected from cosmic variance (Ilbert et al. 2006), it is in excellent
agreement with the COSMOS-30 redshift distribution. This agreement
suggests that the derivation of $\sigma_8(\Omega_{\rm
m}/0.25)^{0.64}=0.785\pm 0.043$ by Fu et al. (2008) is not suffering
from biases due to the photo-z or significant cosmic variance.

The COSMOS-30 redshift distribution can be fit with a parametrization
similar to that used by Fu et al. (2008):
\begin{equation}\label{NZ}                         
n(z) = A \frac{(z^a+z^{ab})}{(z^b+c)}
\end{equation}
where A is the normalization factor and $a$, $b$ and $c$ are free parameters.
The best fit parameters $a$, $b$ and $c$ are given in Table~\ref{fu},
as well as the median redshifts. Fig.~\ref{distz_cosmos} shows the
best fit redshift distribution per apparent magnitude bin.  As
expected, the median redshift increases at fainter apparent magnitude,
ranging from $z_{\rm m}=0.66$ at $22<i^+ <22.5$ to $z_{\rm m}=1.06$ at
$24.5<i^+ <25$.

\begin{table*}[htb]
\begin{tabular}{ l l c c c c c } \hline \\
     Sample      &                   &   N   &  median $z_s$  &  median $i^+$  &   $\sigma_{\Delta z /(1+z_{\rm s})}$  &  $\eta$ in \% \\
\\
\hline\\
 zCOSMOS bright  & $17.5<i^+<22.5$   &  4146 &      0.48      &    21.6        &       0.007                           &   0.7       \\ 
 zCOSMOS faint   & $1.5<z_s<3$       &   147 &      2.20      &    24.0        &       0.054                           &   20.4 \\
 MIPS bright     & $17.5<i^+<22.5$   &   186 &      0.68      &    21.7        &       0.009                           &   0.0  \\
 MIPS faint      & $22.5<i^+<24.0$   &   116 &      0.90      &    23.1        &       0.011                           &   0.0   \\
 MIPS very faint & $24.0<i^+<25.0$   &    15 &      1.15      &    24.2        &       0.053                           &   20.0    \\
\hline
\end{tabular}
\caption{Redshift accuracies estimated from the comparison between photometric redshifts and spectroscopic redshifts.}
\label{sigma}
\end{table*}

\section{Summary}

This paper presents a new version of the photometric redshift catalog
for the 2-deg$^2$ COSMOS survey computed with new ground-based NIR
data, deeper IRAC data and a new set of 12 medium bands from the
Subaru Telescope. The COSMOS photometry now includes a total of 30
filters -- from the UV ({\it GALEX}) to the MIR ({\it
Spitzer}-IRAC). The photo-z catalogue derived here contains 607,617
sources at $i^+ <26$.  The 1887 {\it XMM}-COSMOS sources (mainly AGN)
are not included in this catalogue; their photo-z are derived in
Salvato et al. (2008) with similarly good accuracy using a set of
templates for composite AGN/galaxies.

The galaxy photo-z were tested and improved using spectroscopic
redshift samples from the zCOSMOS and Keck surveys. Biases in the
photo-z were removed by iterative calibration of the photometric band
zero-points. As suggested by the data, two different dust extinction
laws were applied specific to the different SED templates. A new
method to take into account the emission lines was implemented using
relations between the UV continuum and the emission line fluxes
associated with star formation activity. The allowance for emission
lines decreased the photo-z dispersion by a factor 2.5.

Based on a comparison between our new values for $z_{\rm p}$ and 4148
measured values of $z_{\rm s}$ from zCOSMOS, we estimate an accuracy
of $\sigma_{\Delta z/(1+z_{\rm s})}=0.007$ for the galaxies brighter
than $i^+ =22.5$. The accuracies measured with the various
spectroscopic samples are summarized in table.\ref{sigma}. We
extrapolate this result to fainter magnitudes using the 1$\sigma$
uncertainties in the photometric redshift probability functions. This
is found to provide reliable uncertainty estimates for the photo-z
technique developed here since these uncertainties agree with the
dispersion in the offsets of photo-z from spec-z for the spectroscopic
sample.  At $z<1.25$, we estimate a photo-z accuracy of
$\sigma_{\Delta z}=0.02$, 0.04, 0.07 for $i^+ \sim 24$, $i^+ \sim 25$,
$i^+ \sim 25.5$. The accuracy is strongly degraded at $i^+>25.5$ and
the exploitation of the COSMOS-30 photo-z at fainter magnitudes should
be done carefully. The accuracy is 3-5 times better than the photo-z
determined for the CFHTLS-DEEP (Ilbert et al. 2006) and the previous
COSMOS photo-z release (Mobasher et al. 2007), and 2 times better than
the accuracy of the photo-z determined for COMBO-17 at $i^+<23$ (Wolf
et al. 2004). Deep NIR ($J$, $K$) and IRAC data were essential to keep
the catastrophic failures low at $z>1.25$. We note that the accuracy
of the COSMOS photo-z could soon be further improved at z $> 1.25$
with the addition of new data currently being obtained in the $Y$ band
(UKIRT), $H$ band (CFHT) and the ULTRA-VISTA survey.

Our photo-z catalogue contains 607,617 sources at $i^+<26$. The
accurate photo-z derived here for this extremely large sample of
galaxies are crucial to scientifically exploit the full legacy value
of the multi-$\lambda$ data sets in the COSMOS field ({\it HST}-ACS,
{\it Spitzer}, {\it GALEX}, VLA, {\it XMM} and {\it Chandra}).

\acknowledgments 
We gratefully acknowledge the contributions of the entire COSMOS
collaboration consisting of more than 100 scientists.  The {\it HST}
COSMOS program was supported through NASA grant HST-GO-09822. French
co-authors acknowledge support from the French Agence National de la
Recheche fund ANR-07-BLAN-0228 (DESIR project). More information on
the COSMOS survey is available at http://www.astro.caltech.edu/cosmos
.  We also greatly appreciate the hospitality provided by the Aspen
Center for Physics where this manuscript was completed.

\clearpage

\end{document}